\documentclass[]{aastex63}
\usepackage{amsmath}
\usepackage{soul}
\usepackage{CJK}
\usepackage{multirow}
\usepackage{comment}
\usepackage{hyperref}

\newcommand{\psj}[1]{Planet. Sci. J.}

\defcitealias{2016ApJ...817L..16S}{S16}
\defcitealias{2017AJ....154..261C}{CK17}
\defcitealias{2017ApJ...847L..22F}{F17}
\received{December 12, 2021}
\revised{June 9, 2022}
\accepted{June 17, 2022}
\submitjournal{ApJ}
\shorttitle{Haziness Trends in Warm Exoplanets}
\shortauthors{Austin H. Dymont et al.}
\graphicspath{{./}{figures/}}
\begin{document}

\title{Cleaning our Hazy Lens: Exploring Trends in Transmission Spectra of Warm Exoplanets}

\correspondingauthor{Xinting Yu}
\email{xintingyu@ucsc.edu}

\author[0000-0003-0998-9452]{Austin H. Dymont}
\affiliation{Department of Physics, University of California, Santa Cruz \\ 1156 High Street, Santa Cruz, California 95064, USA.}

\author[0000-0002-7479-1437]{Xinting Yu\begin{CJK*}{UTF8}{gbsn}
(余馨婷)\end{CJK*}}
\affiliation{Department of Earth and Planetary Sciences, University of California, Santa Cruz \\ 1156 High Street, Santa Cruz, California 95064, USA.}

\author[0000-0003-3290-6758]{Kazumasa Ohno\begin{CJK*}{UTF8}{gbsn}
(大野和正)\end{CJK*}}
\affiliation{Department of Astronomy and Astrophysics, University of California, Santa Cruz \\ 1156 High Street, Santa Cruz, California 95064, USA.}

\author[0000-0002-8706-6963]{Xi Zhang}
\affiliation{Department of Earth and Planetary Sciences, University of California, Santa Cruz \\ 1156 High Street, Santa Cruz, California 95064, USA.}

\author[0000-0002-9843-4354]{Jonathan J. Fortney}
\affiliation{Department of Astronomy and Astrophysics, University of California, Santa Cruz \\ 1156 High Street, Santa Cruz, California 95064, USA.}

\author[0000-0002-5113-8558]{Daniel Thorngren}
\affiliation{Institute for Research on Exoplanets (iREx), Universit\'e de Montr\'eal, Montreal, QC, Canada}

\begin{abstract}
Relatively little is understood about the atmospheric composition of temperate to warm exoplanets (equilibrium temperature $T_{\rm eq}<$ 1000 K), as many of them are found to have uncharacteristically flat transmission spectra. Their flattened spectra are likely due to atmospheric opacity sources such as planet-wide photochemical hazes and condensation clouds. We compile the transmission spectra of 25 warm exoplanets previously observed by the \textit{Hubble Space Telescope} and quantify the haziness of each exoplanet using a normalized amplitude of the water absorption feature ($A_{\rm H}$). By examining the relationships between $A_{\rm H}$ and various planetary and stellar forcing parameters, we endeavor to find correlations of haziness associated with planetary properties. We adopt new statistical correlation tests that are more suitable for the small, non-normally distributed warm exoplanet sample. Our analysis shows that none of the parameters hold statistically significant correlation with $A_{\rm H}$ ($p \le 0.01$) with the addition of new exoplanet data, including the previously identified linear trends between $A_{\rm H}$ and $T_{\rm{eq}}$ or hydrogen-helium envelope mass fraction (f$_{\rm{HHe}}$). This suggests that haziness in warm exoplanets is not simply controlled by any single planetary/stellar parameter. Among all the parameters we investigated, planet gravity ($g_{\rm p}$), atmospheric scale height ($H$), planet density ($\rho_{\rm p}$), orbital eccentricity ($e$), and age of the star ($t_{\rm age}$) hold tentative correlations with $A_{\rm H}$. Specifically, lower $H$, higher $g_{\rm p}$, $\rho_{\rm p}$, $e$, or $t_{\rm age}$ may lead to clearer atmospheres. We still need more observations and laboratory experiments to fully understand the complex physics and chemistry involved in creating hazy warm exoplanets. 
\end{abstract}


\section{Introduction} \label{sec:intro}

In the past few decades, astronomical advances have resulted in various new techniques for exoplanet detection and thousands of exoplanet discoveries \citep[e.g.,][]{2011arXiv1109.2497M,2013ApJ...766...81F, 2014PASP..126..398H, 2015JATIS...1a4003R, 2018ApJS..235...38T, 2020AJ....159..248K}. These exoplanets vary drastically in size, mass, and composition, which give us a large sample size to understand trends of exoplanetary properties \citep[e.g.,][]{2017AJ....154..108J, 2017AJ....154..109F, 2018AJ....156..264F}. The most common types of exoplanets are with radius ($R_{\rm p}$) between Earth-sized and Neptune-sized planets ($1R_{\oplus}< R_p < 4 R_{\oplus}$); henceforth, referred to as ``sub-Neptunes''\footnote{This definition of sub-Neptunes also includes the so-called ``super-Earth" exoplanets, whose sizes are smaller than $2 R_{\oplus}$ \citep{2013ApJ...766...81F}.} \citep{2012ApJS..201...15H, 2013ApJ...766...81F, 2013PNAS..11019273P, 2013ApJ...770...69P}. 

More recently, there have been extensive observational studies on characterizing the atmospheric compositions of these sub-Neptunes using transmission spectroscopy. In particular, water absorption features were discovered in the transmission spectra taken from the Wide Field Camera Three (WFC3) of the \textit{Hubble Space Telescope} (\textit{HST}) for several warm sub-Neptunes
\citep[e.g,][]{2014Natur.513..526F, 2017Sci...356..628W, 2018ApJ...858L...6K, 2019ApJ...887L..14B, 2019NatAs...3.1086T, 2019NatAs...3..813B, 2020arXiv200607444K, 2020AJ....159..239G, 2021AJ....161...44E,2021AJ....161...18M}. Meanwhile, many transmission spectra were found to be flat and featureless \citep[e.g,][]{2014Natur.505...69K, 2014Natur.505...66K, 2014ApJ...794..155K, 2020AJ....159...57L, 2020AJ....160..201C, 2021AJ....161..284M}. Most notably, GJ 1214 b, with 15 transits, was observed to have an almost completely flat spectrum over the WFC3 range of $\sim$1.1--1.7 $\mu$m \citep{2014Natur.505...69K}. Several reasons could lead to such observed flat spectra: 1) an absence of H$_2$O and/or CH$_4$ in these atmospheres; 2) the atmospheres having high mean molecular weight compositions; 3) and opacity sources such as high-altitude condensation clouds or photochemical hazes obscuring the H$_2$O/CH$_4$ spectral features.

It is believed that fine solid/liquid particles suspended in the atmosphere, called aerosols, likely act as the opacity sources that flatten the transmission spectra \citep[e.g.,][]{
Brown01, Fortney05, Howe&Burrows12, 2013ApJ...775...33M, 2013ApJ...775...80F, Charnay+15, 2016Natur.529...59S, Lavvas&Koskinen17, Kawashima&Ikoma18, Gao&Benneke18, Adams+19, Powell+19, Ohno+20a}. Condensation clouds and photochemical hazes are two common types of aerosols that can mute spectral features \citep[for a recent review, see][]{Gao+21}. Laboratory experiments suggest a likely ubiquity of photochemical hazes in warm exoplanet atmospheres with equilibrium temperature $T_{\rm eq}\leq800$ K \citep[][]{Horst17, 2018NatAs...2..303H, 2018ApJ...856L...3H, 2018AJ....156...38H, 2019ECS.....3...39H, 2020NatAs...4..986H, 2020PSJ.....1...51H, 2020PSJ.....1...17M,
2021PSJ.....2....2V,
2021NatAs...5..822Y}. In addition, photochemical models suggest haze formation is likely promoted in temperate to warm exoplanet atmospheres with $T_{\rm eq}\le1000$ K \citep[][]{2013ApJ...775...80F, Morley+15, 2019ApJ...884...98K}. Aerosols are also suggested to impact the thermal structure of exoplanet atmospheres \citep[e.g.,][]{Heng+12,Morley+15,Lavvas&Aufaux21}, which potentially affects atmospheric chemical compositions \citep{Molaverdikhani+20} and dynamics \citep{Roman&Rauscher17}. Aerosols can also significantly inflate the observable transit radius from that expected for clear atmospheres, complicating the interpretation of the observed radius \citep{Lammer+16,Wang&Dai19,Kawashima+19,2020ApJ...890...93G,Ohno&Tanaka21}. Understanding the effect of these opacity sources are vital to unveil the atmospheric properties, such as composition and dynamics, of sub-Neptunes.

To unveil the nature of cloudy/hazy exoplanetary atmospheres, it is vital to understand how the atmospheric cloudiness/haziness depends on planetary properties. \citet{2016Natur.529...59S} and \citet{2016ApJ...817L..16S} (henceforth \citetalias{2016ApJ...817L..16S}) first developed a metric to cross-compare the observed H$_2$O absorption feature's amplitude between different exoplanet atmospheres, the so-called ``water amplitude" ($A_{\rm H}$), using transmission spectroscopy data. 
The ``water" amplitude measures the flatness of the spectra observed by {\it HST}/WFC3 based on the difference of transit depth between the H$_2$O absorption band (wavelength $\lambda\sim1.4\ \mu$m) and the baseline J-band ($\lambda\sim1.25\ \mu$m).
The water amplitude is an estimated predictor for cloudiness/haziness for a given exoplanet, given that clouds and hazes are the main cause of the featureless spectra. \citetalias{2016ApJ...817L..16S} and \citet{2017ApJ...847L..22F} (henceforth \citetalias{2017ApJ...847L..22F}) set out to discover empirical trends for hot-Jupiters and some sub-Neptunes, and established a linear trend between the cloudiness/haziness versus $T_{\rm eq}$ \citepalias[][]{2016ApJ...817L..16S, 2017ApJ...847L..22F}. 
\citet{2017AJ....154..261C} (henceforth \citetalias{2017AJ....154..261C}) found a similar linear trend between $A_{\rm H}$ vs $T_{\rm eq}$ and the predicted hydrogen-helium mass fraction ($f_{\rm HHe}$) for warm sub-Neptunes, though their sample size is considerably smaller ($N=6$ for \citetalias{2017AJ....154..261C}; $N=14$ for  \citetalias{2016ApJ...817L..16S}; and $N=34$ for \citetalias{2017ApJ...847L..22F}). Based on the aforementioned trend, \citet{2020NatAs...4..951G} suggested that condensed silicate clouds are likely the dominant opacity sources in hot exoplanets ($T_{\rm eq}>950\ K$) while photochemical hazes likely dominate atmospheric opacity at lower temperature exoplanets ($T_{\rm eq}<950\ K$). Recently, there has also been some efforts looking into trends of haziness/cloudiness for hot-Jupiters using the thermal emission spectra data \citep{Mansfield+21}.

In this study, we aim to expand upon previous works on characterizing the cloudiness/haziness of exoplanets, focusing on temperate to warm exoplanets with $T_{\rm{eq}}$ less than 1000 K, where photochemical hazes are likely the dominant opacity source \citep{2020NatAs...4..951G}. Compared to the previous work of \citetalias{2017AJ....154..261C}, we include new observations, especially for colder exoplanets. We also investigate trends for more planetary/stellar parameters. However, even with the increased sample size, our exoplanet sample are still small (25 exoplanets) and shows non-normal distribution of data. Thus, we choose to use non-parametric statistical correlation tests, similarly to \citetalias{2017ApJ...847L..22F}, such as the Spearman's rho and Kendall's tau  \citep[see][]{2012msma.book.....F}, instead of the parametric tests (e.g., Pearson correlation coefficient or Chi-squared goodness of fit test) used in \citetalias{2016ApJ...817L..16S} and \citetalias{ 2017AJ....154..261C}. We investigate whether previously established trends hold true, and explore whether we can establish new trends between haziness and planetary/stellar parameters. 
These trends, if any exist, will help us better understand the physics of haze formation and removal on exoplanets and assess under which circumstances we are more likely to see clear versus hazy exoplanets for future observations.

The organization of this paper is as follows.
In Section \ref{sec:method}, we detail our water amplitude calculations. We also describe the calculations or estimations of all relevant planetary and stellar parameters. In Section \ref{sec:a_analysis}, we show our test statistics results of stellar/planetary parameters versus water amplitudes and compare them with previous results \citepalias{2016ApJ...817L..16S, 2017AJ....154..261C, 2017ApJ...847L..22F}. In Section \ref{sec:b_analysis}, we attempt to find analytical dependencies between the water amplitude and a combination of planetary parameters that are motivated by the microphysics of haze formation. In Section \ref{sec:discussion}, we discuss the validity and caveats of our results.

\section{Methods} \label{sec:method}

\subsection{Exoplanet Targets}
We select exoplanet targets mainly based on their equilibrium temperature ($T_{\rm eq}<1000$ K), where we assumed a planetary Bond albedo (A$_b$) of $0.3$, regardless of their masses or sizes. Our targets are mostly sub-Neptunes ($R_{\rm p} < 4 R_{\oplus}$), but it also includes a few super-puffs (large radii $R_{\rm p} > 4 R_{\oplus}$ but similar masses $M_{\rm p} \la 12 M_{\oplus}$ as sub-Neptunes), Neptune to super-Neptune sized (4$ R_{\oplus}
< R_p < 13 R_{\oplus})$, and terrestrial-sized exoplanets ($R_{\rm p} < 2 R_{\oplus}$). The targets are summarized in Table \ref{tab:param}. Overall, our exoplanet targets have radii ranging from $\sim$1 to 13 $R_{\oplus}$, masses range from $\sim$1 to 185 $M_{\oplus}$ (or $\sim$0.6 $M_J$), and $T_{\rm eq}$ range from $\sim$200 to 1000 K. Even though all the TRAPPIST-1 system planets \citep[][]{2017Natur.542..456G} fit in our $T_{\rm eq}$ criterion, we do not include those with transmission spectra data in our present analysis due to large uncertainties in their transmission spectra \citep{2016Natur.537...69D,2018NatAs...2..214D,2018AJ....156..178Z}.

\subsection{Definition of Water Amplitude}\label{sec:water_amplitude}

In previous works \citepalias{2016ApJ...817L..16S, 2017AJ....154..261C}, the common idea in the definition of the water amplitude metric involves calculating the difference of transit depths at two wavelengths where a spectral feature is expected and not. The transit depth ($D$) measured from transmission spectroscopy is a function of wavelength ($\lambda$):
 \begin{equation}\label{eq:D}
 D(\lambda) = \frac{R_{\rm p}(\lambda)^2}{R_*^2},
 \end{equation}
where $R_{\rm p}(\lambda)$ is the transit planetary radius as a function of wavelength and $R_*$ is the stellar radius.

Specifically, here we estimate the haziness of the transmission spectrum by normalizing the amplitude difference between the H$_2$O absorption feature around 1.4 $\mu$m and the J-band baseline around 1.25 $\mu$m. The normalized transit radius difference between these two wavelengths establishes our water amplitude metric. 

To calculate the water amplitude from transmission spectra, we first compute the inverse-variance weighted average of the transit depths around 1.25 $\mu$m from $\sim$1.22-1.3 $\mu$m, and around 1.4 $\mu$m from $\sim$1.36-1.44 $\mu$m, depending on the spectral binning conventions used between various sources (adapted from \citetalias{2016ApJ...817L..16S}). Next, we compute $R_{\rm p}(\lambda)$ from Equation \eqref{eq:D}, where we substitute in the averaged transit depth ($D_{\rm avg}$) (adapted from \citetalias{2017AJ....154..261C}). Taking the difference of transit radii at $1.4$ and $1.25~{\rm \mu m}$, we define the water amplitude metric as
\begin{equation}\label{eq:Ah}
    A_H = \frac{R_*}{H}\left(\sqrt{D_{\rm avg}(1.4\ \mu m)} - \sqrt{D_{\rm avg}(1.25\ \mu m)} \right),
\end{equation}
where $H=k_{\rm B}{T_{\rm eq}}/{\mu}g_{\rm p}$ is the atmospheric scale height, $k_B$ is the Boltzmann constant, $\mu$ is the atmospheric mean molecular mass, $g_{\rm p}$ is the surface gravity of the planet. To compute $H$, we assume solar composition atmospheres with mean molecular weight of $2.3~{\rm amu}$ following \citetalias{2017AJ....154..261C}, since most of the planets are large enough to have H$_2$-He dominated atmospheres. We also attempted to compute $H$ by assigning a different mean molecular weight for each target using the solar system mass-metallicity relationship in \cite{2017Sci...356..628W} and metallicity-mean molecular weight relationship in \cite{2011ApJ...733....2N}, see discussion in Section 5.2.2. We calculated $T_{\rm eq}$ from the stellar effective temperature ($T_{\text{eff}*}$), radius ($R_*$), planet orbital semi-major axis ($a$), and Bond albedo ($A_{\rm B}$) by the relationship of $T_{\rm eq} = T_{\text{eff}*}(1-A_b)^{1/4}\sqrt{{R_*}/{2a}}$. We choose a Bond albedo of 0.3 because Earth, Mars, Titan, and the giant planets in the Solar System all have Bond albedos around 0.2-0.4 \citep{2001plsc.book.....D}, as well as some hot-Jupiters \citep{2005ApJ...626..523C,2021NatAs...5.1001H}. We will discuss the caveats regarding these assumptions in Section \ref{sec:discussion}. Our method takes a very similar approach compared to \citetalias{2016ApJ...817L..16S} and \citetalias{2017AJ....154..261C} and is a simple way of estimating the water amplitude of exoplanets from available observational data. We will discuss the consistency of the results obtained from our method with previous studies in Section \ref{sec:discussion}.

\begin{figure}
    \centering
    \includegraphics[scale=0.5]{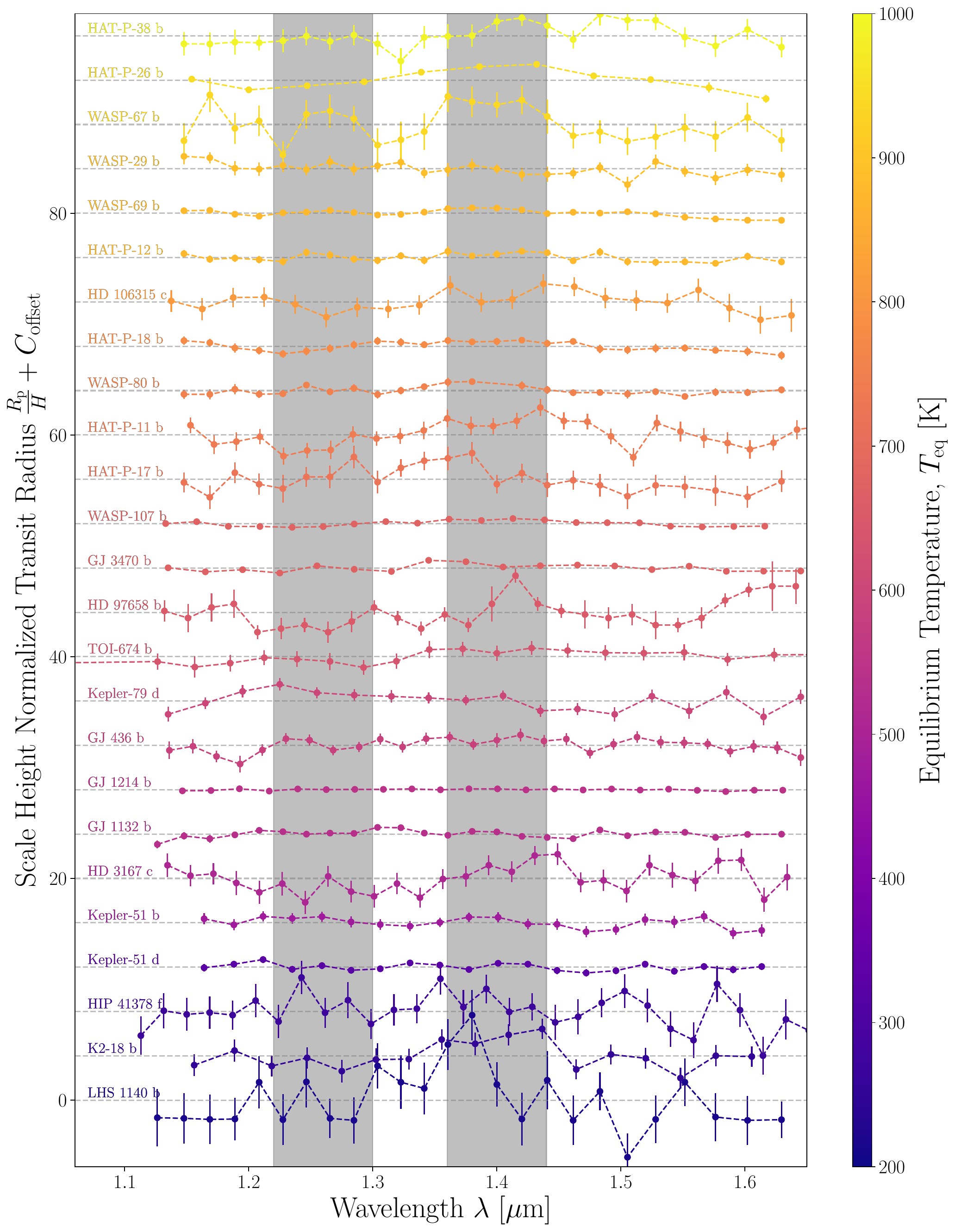}
    \caption{Transmission spectra data (points) connected linearly between each point for clarity (dashes) for our 25 exoplanet targets. We normalize the observed transit radius of each target by the atmospheric scale height ($H$) and offset for clarity. The color of the data points shows the equilibrium temperature ($T_{\rm eq}$) for each target.}
    \label{fig:spectra}
\end{figure}

In Figure \ref{fig:spectra}, we present our adopted transmission spectra for the 25 targets observed with \textit{HST}/WFC3 ordered by $T_{\rm eq}$ (from warm to cool). We select existing reduced transmission spectra data from the literature. For some exoplanets with multiple published reduced data, we adopt the ones that have the most transits or use more up-to-date stellar parameters. We summarize the calculated water amplitudes and our adopted transmission spectra data references for all our exoplanet targets in Table \ref{tab:param}.

Before we search for trends between our water amplitude metric and the various stellar/planetary parameters, we clarify that our definition of water amplitude depends only on atmospheric properties and does not have an explicit dependence on the planetary/stellar parameters. To this end, we analyse our definition of $A_{\rm H}$ using an analytical model of the transmission spectrum. \citet{2017MNRAS.470.2972H} provided an approximate solution for $R_{\rm p}(\lambda)$ as,
\begin{equation}\label{eq:Rp}
    R_{\rm p}(\lambda)\approx R_{\rm 0}+H\left[ \gamma+\ln{\left(\frac{P_{\rm 0}\kappa(\lambda)}{g}\sqrt{\frac{2\pi R_{\rm 0}}{H}} \right)} \right],
\end{equation}
where $\gamma\approx0.57721$ is the Euler-Mascheroni constant, $R_{\rm 0}$ is the reference planetary transit radius, $\kappa(\lambda)$ is the atmospheric opacity---extinction cross sections per mass---as a function of wavelength, and $P_{\rm 0}$ is the pressure at $R_{\rm 0}$. Equation \eqref{eq:Rp} is valid for isothermal atmospheres with vertically constant gravity and opacity \citep[for vertically varying opacity, see][]{Ohno&Kawashima20}. 
Inserting Equation \eqref{eq:Rp} into Equation \eqref{eq:Ah}, we can express the water amplitude as:
\begin{eqnarray}\label{eq:Ah2}
    \nonumber
    A_{\rm H}&\approx&\frac{R_{\rm *}}{H}\frac{H}{R_{\rm *}}\left[ \ln{\left(\frac{P_{\rm 0}\kappa(1.4~{\rm \mu m})}{g}\sqrt{\frac{2\pi R_{\rm 0}}{H}} \right)}-\ln{\left(\frac{P_{\rm 0}\kappa(1.25~{\rm \mu m})}{g}\sqrt{\frac{2\pi R_{\rm 0}}{H}} \right)}\right]\\
    &=&\ln{\left(\frac{\kappa(1.4~{\rm \mu m})}{\kappa(1.25~{\rm \mu m})}\right)}.
\end{eqnarray}
Equation \eqref{eq:Ah2} demonstrates that our water amplitude metric does not have any explicit dependence on any specific planetary/stellar parameters. The water amplitude is a measure of the ratio of atmospheric opacity between the two wavelengths. For a cloud/haze free atmosphere with water vapor, $A_{\rm H}$ should be around 7 \citepalias{2017AJ....154..261C}. If the water amplitude depends on planetary/stellar parameters, this indicates that atmospheric opacities have implicit dependencies on stellar/planetary parameters. We will explore the possible implicit dependencies of the opacities for hazy atmospheres using a simple but physically motivated haze model in Section \ref{sec:b_analysis}.

\begin{deluxetable}{l|Cl|C@{$\pm$}Cl|C@{$\pm$}Cl|C@{$\pm$}C|C@{$\pm$}C|C@{$\pm$}C|C@{$\pm$}C|c}
\caption{\large{Summary of Parameters}\label{tab:param}}
\tablehead{
\colhead{Planet Name} & \multicolumn{2}{c}{$t_{\rm age}$}  & \multicolumn{3}{c}{$M_{\rm p}$ }  & \multicolumn{3}{c}{$R_{p}$ } & \multicolumn{2}{c}{$T_{\rm eq}$} & \multicolumn{2}{c}{$g_{\rm p}$} & \multicolumn{2}{c}{$H$} & \multicolumn{2}{c}{$A_{\rm H}$} &\colhead{WFC3}\\ \colhead{} & \multicolumn{2}{c}{[Gyr]} & \multicolumn{3}{c}{[$M_{\oplus}$]} & \multicolumn{3}{c}{[$R_{\oplus}$]} & \multicolumn{2}{c}{[K]} & \multicolumn{2}{c}{[m$^2$/s]} & \multicolumn{2}{c}{[km]} & \multicolumn{2}{c}{[$H$]} & \colhead{Ref(s).}
}
\decimals
\tablecolumns{8}
\startdata
GJ 436 b    & 6    ^{+4}_{-5}   &[1] & 25.4  & 2.1  &[2]  & 4.10    & 0.16 &[2]  & 586    & 10    &     14.81 &   1.68 & 143    & 16.45   & 0.43 & 0.34 & [a] \\[0.5ex]
GJ 1132 b   & 9\pm4  & [3] & 1.66  & 0.23 &[4]  & 1.13   & 0.056&[5]  & 535    & 23    &     12.74 &   2.17 & 152    & 26.67   & -0.31 & 0.19 & [b,c,d]  \\[0.5ex]
GJ 1214 b   & 6.5 \pm 3.5  &[6] & 6.26  & 0.91 &[7]  & 2.80    & 0.24 &[7]  & 544   & 3     &      7.82 &   1.76 & 252    & 56.56   & 0.02 & 0.05 & [e] \\[0.5ex]
GJ 3470 b   & 1.8 \pm 1.2  &[8] & 12.58 & 1.31 &[9]  & 3.88   & 0.32 &[10] & 670    & 10    &      8.19 &   1.60 & 296    & 57.89   & 0.42 & 0.19 & [f]\\[0.5ex]
HAT-P-11 b  & 6.5  ^{+5.9}_{-4.1}   &[11]& 23.4  & 1.5  &[12] & 4.36   & 0.06 &[12] & 740    & 8     &     12.06 &   0.84 & 222    & 15.64   & 2.46 & 0.54 & [g] \\[0.5ex]
HAT-P-12 b  & 2.5 \pm 2   & [13]& 67.1  & 3.8  &[13] & 10.7   & 0.3  &[13] & 877    & 11    &      5.69 &   0.47 & 557    & 46.82   & 0.09 & 0.30 & [h]\\[0.5ex]
HAT-P-17 b  & 7.8  \pm 3.3  &[14]& 184  & 19.0  &[15] & 11.8   & 0.4  &[15] & 724    & 16    &     13.04 &   1.67 & 201    & 26.17   & 0.76 & 0.87 & [h]\\[0.5ex]
HAT-P-18 b  & 12.4 ^{+4.4}_{-6.4}  &[16]& 62.6 & 4.1   &[16] & 11.2   & 0.6 &[16] & 776     & 13    &      4.93 &   0.61 & 569    & 70.93   & 0.88 & 0.31 & [h] \\[0.5ex]
HAT-P-26 b  & 9  ^{+3}_{-4.9}  &[17]& 18.75 & 2.23 &[17] & 6.03   & 0.75 &[18] & 946    & 16    &      5.06 &   1.39 & 677    & 186.8   & 1.84 & 0.35 & [i]\\[0.5ex]
HAT-P-38 b  & 10.1 ^{+3.9}_{-4.8} &[14]& 84.9 & 6.4   &[19] & 9.247  & 1.031&[19] & 988    & 19    &      9.72 &   2.29 & 367    & 86.69   & 1.06 & 0.82 & [h] \\[0.5ex]
HD 3167 c   & 7.8  \pm 4.3  &[20]& 9.8 & 1.3    &[20] & 3.01   & 0.42 &[20] & 512    & 6     &     10.60 &   3.27 & 174    & 53.95   & 1.82 & 0.65 & 
[j] \\[0.5ex]
HD 97658 b  & 6.1  \pm 0.7  &[21]& 7.81 & 0.55  &[22] & 2.303  & 0.110 &[22] & 650   & 15    &     14.43 &   1.71 & 163    & 19.7    & 1.44 & 0.47 & [k,l] \\[0.5ex]
HD 106315 c & 4.48 \pm 0.96 &[23]& 15.2 & 3.7   &[23] & 4.35   & 0.23 &[23] & 810    & 6     &      7.87 &   2.09 & 372    & 98.82   & 1.65 & 0.67 &
[m]\\[0.5ex]
HIP 41378 f & 3.1  \pm 0.4 & [24]  & 12.3    & 3.1  & [24]    & 9.2    & 0.1 & [24]  & 269    & 2     &   1.42 &   0.36& 683    & 172.8   & 0.35 & 0.45 &[n] \\[0.5ex]
Kepler-51 b & 0.5  \pm 0.25 &[25]& 3.69 & 1.86  &[25] & 6.89   & 0.14 &[25] & 477    & 5     &      0.76 &   0.39 & 2270   & 1146    & 0.00 & 0.33 & [o] \\[0.5ex]
Kepler-51 d & 0.5  \pm 0.25 &[25]& 5.70  & 1.12  &[25] & 9.46   & 0.16 &[25] & 328   & 4     &      0.62 &   0.12 & 1900   & 379.7   & 0.16 & 0.21 & [o] \\[0.5ex]
Kepler-79 d & 1.3  ^{+1}_{-0.4} &[26]& 5.3 & 0.9    &[26] & 7.2   & 0.2  &[26] & 597     & 6     &      1.02 &   0.18 & 2120   & 380.1   & -0.93 & 0.41 & [p] \\[0.5ex]
K2-18 b     & 2.4  \pm 0.6  &[27]& 8.63 & 1.35  &[28] & 2.610   & 0.087&[29] & 255   & 3     &     12.41 &   2.11 & 74.2   & 12.65   & 2.36 & 0.74 & [q,r] \\[0.5ex]
LHS 1140 b  & 9   \pm  4    &[30]& 6.98 & 0.89  &[31] & 1.727  & 0.032&[31] & 214    & 3     &     22.93 &   3.04 & 33.8   & 4.508   & 3.07 & 1.38 & [s]\\[0.5ex]
TOI-674 b   & 5.5^{+2.9}_{-1.9} & [32] & 23.6    & 3.3    & [32] & 5.25   & 0.17  & [32] & 635    & 10    &   8.39 &   1.29& 274    & 42.42   & 1.19 & 0.53 & [t] \\[0.5ex]
WASP-29 b   & 10.5 \pm 3.5    &[14]& 73 & 16     &[15] & 8.6    & 0.8  &[15] & 890     & 28    &      9.61 &   2.73 & 334    & 95.38   & -0.17 & 0.47 & [h] \\[0.5ex]
WASP-67 b   & 12.6 ^{+1}_{-4.2}  &[14]& 137 & 29    &[15] & 12.9   & 1.2  &[15] & 939     & 18    &      8.06 &   2.29 & 421    & 119.7   & 2.33 & 1.16 & [h] \\[0.5ex]
WASP-69 b   & 1   \pm  1    &[14]& 92 & 10     &[15] & 12.4   & 0.4  &[15] & 878    & 9     &       5.83 &   0.74 & 544    & 68.78   & 0.35 & 0.17 & [h]\\[0.5ex]
WASP-80 b   & 7    \pm 7    &[14]& 171 & 11    &[15] & 11.2   & 0.3 &[15] & 754     & 17    &      13.36 &   1.18 & 204    & 18.62   & 0.61 & 0.29 & [h] \\[0.5ex]
WASP-107 b  & 8.3  \pm 4.3  &[33]& 30.5 & 1.7   &[34] & 10.39  & 0.33 &[35] & 676   & 11    &      2.77 &   0.23 & 884    & 75.81   & 0.59 & 0.10 & [v]\\[0.5ex]
\enddata
\tablecomments{For all parameter values and references, see Appendix \ref{ap:A}. We use the first reference listed in our calculation of $A_{\rm H}$. We select the reference based upon it being the most up-to-date published result and consistent/similar method of data reduction.}
\tablerefs{
[1] \citet{2007ApJ...671L..65T}; [2] \citet{2014AA...572A..73L}
[3] \citet{2015Natur.527..204B}; [4] \citet{2018AA...618A.142B}; [5] \citet{2017AJ....154..142D}; 
[6] \citet{2009Natur.462..891C}; [7] \citet{2013AA...551A..48A};
[8] \citet{2020AA...638A..61P}; [9] \citet{2019AJ....157...97K}; [10] \citet{2014MNRAS.443.1810B};
[11] \citet{2010ApJ...710.1724B}; [12] \citet{2018AJ....155..255Y};
[13] \citet{2009ApJ...706..785H}; 
[14] \citet{2017AA...602A.107B}; [15] \citet{2017AJ....153..136S};
[16] \citet{2011ApJ...726...52H};
[17] \citet{2011ApJ...728..138H}; [18] \citet{2019AA...628A.116V};
[19] \citet{2012PASJ...64...97S}; 
[20] \citet{2017AJ....154..122C};
[21] \citet{2011arXiv1109.2549H}; [22] \citet{2020AJ....159..239G}
[23] \citet{2017AA...608A..25B};
[24] \citet{2019arXiv191107355S}
[25] \citet{2020AJ....159...57L};
[26] \citet{2020AJ....160..201C}; 
[27] \citet{2019RNAAS...3..189G}; [28] \citet{2019AA...621A..49C}; [29] \citet{2019NatAs...3..813B};
[30] \citet{2017Natur.544..333D}; [31]\citet{2019AJ....157...32M}; 
[32] \citet{2021AA...653A..60M}
[33] \citet{2017MNRAS.469.1622M}; [34] \citet{2021AJ....161...70P}; [35] \citet{2017AJ....153..205D}; \\
WFC3 Observations/Data --- 
[a] \citet{2014Natur.505...66K}; 
[b] \citet{2021AJ....161..284M}; [c] \citet{2021arXiv210510487L}; [d] \citet{2021AJ....161..213S};
[e] \citet{2014Natur.505...69K}; 
[f] \citet{2019NatAs...3..813B};
[g] \citet{2014Natur.513..526F};
[h] \citet{2018AJ....155..156T};
[i] \citet{2017Sci...356..628W} 
[j] \citet{2021AJ....161...18M};
[k] \citet{2020AJ....159..239G};
[l] \citet{2014ApJ...794..155K};
[m] \citet{2020arXiv200607444K};
[n] \citet{2022ApJ...927L...5A} 
[o] \citet{2020AJ....159...57L};
[p] \citet{2020AJ....160..201C};
[q] \citet{2019ApJ...887L..14B}; [r] \citet{2019NatAs...3.1086T};
[s] \citet{2021AJ....161...44E};
[t] \citet{2022arXiv220104197B}; 
[u] \citet{2018ApJ...858L...6K};
}
\end{deluxetable}

\subsection{Stellar Forcing and Planetary Parameters}\label{sec:param}

In this section, we summarize all the stellar forcing and planetary parameters that we used to show trends of the water amplitude metric for the exoplanet targets.

For the properties related to stellar forcing, we collected from the literature the spectral type, age ($t_{\rm age}$ [Gyr]), mass ($M_*$ [$M_{\odot}$]), radius ($R_*$ [$R_{s}]$]), effective temperature ($T_\text{eff*}$ [K]), orbital semi-major axis ($a$ [AU]), metallicity ($Z_{\rm [Fe/H]}$ [dex]), eccentricity ($e$), and stellar rotation period ($P_{\rm rot}$ [day]). We also calculated the density ($\rho_*$ [$\rho_{\odot}$]), surface gravity ($g_*$ [m s$^{-2}$]), and bolometric luminosity ($L_{\rm bol}$ [erg s$^{-1}$]) for the stars. For the near-UV (NUV, 170--320 nm) flux density ($F_{\rm NUV}$ [erg cm$^{-2}$ s$^{-1}$]), far-UV (FUV, 91.2--170 nm) flux density ($F_{\rm FUV}$ [erg cm$^{-2}$ s$^{-1}$]), and X-rays plus extreme-UV (or the so-called XUV, 0.5--91.1 nm) flux density ($F_{\rm XUV}$ [erg cm$^{-2}$ s$^{-1}$]), we use a combination of literature and estimated values. All the stellar forcing parameters are summarized in Tables \ref{tab:s-param} and \ref{tab:sy-param} in Appendix \ref{ap:A}. 

For HD 97658b, GJ 436b, GJ 1214b, we used their measured and reconstructed FUV, NUV, and XUV flux luminosities from the MUSCLES TREASURY Survey reported by \citet{2016ApJ...820...89F}, and converted them into a flux density at their orbital semi-major axis, $a$. We are also able to estimate the FUV, NUV, XUV flux densities for three additional planets given their measured flux spectra taken with \textit{HST}-COS G130M mode: HAT-P-11b, GJ 3470b, and WASP-69b. For these three targets, we scaled their UV fluxes using a combination of the reconstructed MUSCLES UV flux data and the \textit{HST}-COS data \citep{2018ApJS..239...16F}. We first calculate the ratios for the integrated fluxes at Si IV ($\sim$1400 $\AA$) and NV ($\sim$1240 $\AA$) emission lines between the target star and a MUSCLES star using the \textit{HST}-COS data. We then scaled the NUV, FUV, XUV fluxes of our target stars at their corresponding wavelengths by multiplying the ratio to the reconstructed MUSCLES star data. Finally, we calculate the converted stellar UV fluxes into flux densities at the planets orbital semi-major axis. We scaled HAT-P-11 (stellar type K4) and WASP-69 (K5) using the MUSCLES star HD 85512 (K6) as a proxy, and GJ-3470 (M1.5) using the MUSCLES star GJ-832 (M1.5) as a proxy (K. France, private communications). We find the emission line flux ratio to be $\sim0.9-2.0$ between HD-85512 and Hat-P-11, $\sim3.3-4.6$ between HD-85512 and WASP-69, and $\sim1.1-1.8$ between GJ 832 and GJ 3470.

For the rest of the planets without directly measured UV fluxes, we estimated their $F_{\rm NUV}$ and $F_{\rm FUV}$ using the following relationships, which are fitted using the reported flux values of 13 FGK stars from \cite{2013ApJ...763..149F, 2016ApJ...820...89F}:

\begin{equation}\label{eq:NUV}
\begin{aligned}
    log_{10}(F_{NUV}) &= 9.63\times10^{-4}T_{\text{eff}*}-7.16 + log_{10}\left(\frac{L_{\rm bol}}{4{\pi}a^2}\right), \\
    log_{10}(F_{FUV}) &= -9.85\times10^{-4}T_{\text{eff}*} + 2.8 + log_{10}(F_{NUV}). \\
\end{aligned}
\end{equation}

We estimated the XUV flux densities ($F_{\rm XUV}$) for these planets using a combination of published relationships in the literature. We can use the age of their parent stars ($t_{\rm age}$) to estimate the X-ray flux densities (F$_X$, $\sim$0.5--10 nm) of the star. In particular, we use the relationship in \citet{2016ApJ...821...81G} for M-type stars, the relationship in \citet{2005ApJ...622..680R} for G-type stars, and \citet{2010AA...511L...8S} for F- and K-type stars. To estimate the EUV flux densities ($F_{\rm EUV}$, $\sim$10--91.2 nm) for G-type stars, we use the relationship in \citet{2005ApJ...622..680R} between $t_{\rm age}$ and $F_{\rm EUV}$. To estimate $F_{\rm EUV}$ for M-, F- and K- stars, we can use Lyman alpha flux densities (F$_{Ly\alpha}$) to estimate their $F_{\rm EUV}$ using relationships given in \citet{2014ApJ...780...61L}. We estimate F$_{Ly\alpha}$ using the fifth method described in \citet{2013ApJ...766...69L}. Combining the estimated $F_{\rm X}$ ($\sim$0.5--10 nm) and F$_{\rm EUV}$ fluxes ($\sim$10--91.2 nm), we get the XUV flux densities ($\sim$0.5--91.2 nm) for these planets.

\begin{deluxetable}{ccccccc}
\caption{\large{XUV Flux Estimation Summary}\label{tab:XUV}}
\tablehead{
\colhead{Star Type} & \colhead{$F_{\rm X}$ (1 AU)} & \colhead{Ref.} & \colhead{$F_{\rm Ly\alpha}$ (1 AU)} & \colhead{Ref.} & \colhead{$F_{\rm EUV}$ (1 AU)} & \colhead{Ref.} 
}
\decimals
\tablecolumns{8}
\startdata
 M-  & $1.3\times10^{0.411-1.424(\log(t_{\rm age}))}$ & [1] & Table 5 & [2] & Table 5 & [3]\\
G- & Table 5 & [4] & ... & ... & Table 5 & [4] \\
F- and K- & $6.72t_{age}^{-1.55}$ & [5] & Table 5 & [2] & Table 5 & [3]\\
\enddata
\tablecomments{All fluxes are normalized fluxes at 1 AU and have units of erg s$^{-1}$ cm$^{-1}$, $t_{\rm age}$ has a unit of Gyr. For plotting and reference proposes, we normalize from 1 AU to each planets $a$ by $F(a) =  F(1\ {\rm AU})/a^2$.}
\tablerefs{[1] \citet{2016ApJ...821...81G}; [2] \citet{2013ApJ...766...69L}; [3] \citet{2014ApJ...780...61L}; [4] \citet{2005ApJ...622..680R}; [5] \citet{2010AA...511L...8S}}
\end{deluxetable}

For each exoplanet, we collected from the literature the mass ($M_{\rm p}$ [$M_{\oplus}$]) and radius ($R_{p}$ [$R_{\oplus}$]). We calculated the equilibrium temperature ($T_{\rm eq}$ [K]), density ($\rho_{\rm p}$ [$\rho_\oplus$]), surface gravity ($g_{\rm p}$ [m s$^{-2}$]), atmospheric scale height ($H$ [m]), and bulk hydrogen-helium (H$_2$-He) mass fraction ($f_{\rm HHe}$ [\%]) of the planets.  All the intrinsic planetary parameters are summarized in Table \ref{tab:p-param} in Appendix \ref{ap:A}.

For most of our planets that are of about a Neptune mass or smaller ($M_{\rm p}<20M_{\oplus}$), we estimated $f_{\rm HHe}$ from \citet{2014ApJ...792....1L}, either using the calculated values from their Table 7 or interpolation from their Table 1-6. For planets above $20M_{\oplus}$, we calculated $f_{\rm HHe}$ following the method described in \citet{2019ApJ...874L..31T}.

\subsection{Statistics}

In this section, we explain the statistical tests we will use to determine the correlation between $A_{\rm H}$ and various stellar forcing/planetary parameters. Because of our small sample size and given that we found most parameters of our exoplanet sample to be non-normally distributed, we choose to use only non-parametric statistical tests in this study.

The first statistical test we use is the Spearman's rho ($\rho_{s}$), which tests for monotonically increasing (or decreasing) behavior between a parameter and $A_{\rm H}$. We also use the Kendall's tau ($\tau_{k}$) which tests for independence of the parameter against $A_{\rm H}$. Both tests are similar in their interpretation to the Pearson's correlation coefficient ($r$) but they do not indicate a particular relationship between the two variables (for example, linear correlation or log-linear correlation). Both tests yield values ranging from $+1$ to $-1$, with zero indicating no correlation or dependency and $\pm1$ indicating completely monotonically increasing (or decreasing) correlation or complete dependency between the data. Given a significant correlation from these two tests, additional tests and fitting would be required to determine the functional form of the trend (e.g., linear). No significant correlation from these tests indicates no trend exists. To determine the significance of these tests, we calculate the associated `p-value' ($p_{\rho}$ for Spearman's rho and $p_{\tau}$ for Kendall's tau) for each of the two tests. The p-values determine the probability of random, uncorrelated data creating a certain value for a given test statistic. Lower p-values reduce concerns of accidental/false correlation between uncorrelated data. In this work, we adopt a threshold of $p\le0.01$ for a statistical correlation to be significant, based on guidance from the American Statistical Association \citep[ASA,][]{ASA}. We note that our small sample size does reduce the reliability of the p-values. More details about the two tests can be found in \citet[][]{2012msma.book.....F}. 

\section{Statistical Analysis}\label{sec:a_analysis}
Table \ref{tab:linearstats} details the test statistics of every stellar forcing and planetary parameter we investigated against water amplitude, $A_{\rm H}$. The Spearman's rho and Kendall's tau tests typically returns very similar results for each investigated parameter. For both tests, none of the parameters hold statistically significant correlations with $A_{\rm H}$ ($p<0.01$). However, some parameters hold tentative correlations over the others. The atmospheric planetary surface gravity ($g_{\rm p}$), scale height ($H$), and planet density ($\rho_{\rm p}$) hold tentative correlations among all the tested parameters ($p\le0.04$). Tentative correlations between $A_{\rm H}$ versus orbital eccentricity ($e$) and stellar age ($t_{\rm age}$) may exist but are plagued with large errors on the parameter values ($p\le0.09$). All the other parameters are obviously not monotonically correlated with $A_{\rm H}$ ($0.31\le p\le0.82$). With the increased amount of water amplitude data (25 in this study versus 6 in \citetalias{2017AJ....154..261C}), previously observed linear trends in \citetalias{2017AJ....154..261C} with equilibrium temperature ($T_{\rm eq}$) and bulk hydrogen-helium fraction ($f_{\rm HHe}$) no longer hold. We describe and discuss in the following sections in detail the correlations for the major stellar forcing and planetary parameters. In addition, we would like to note that, for all the exoplanet targets we investigated, their computed water amplitudes $A_{\rm H}$ are much lower ($A_{\rm H}\la4$) than expected for cloud/haze free atmosphere ($A_{\rm H}\sim7$, \citetalias{2017AJ....154..261C}). This demonstrates that clouds and hazes are indeed ubiquitous in temperate to warm exoplanet atmospheres with $T_{\rm eq}<1000$ K. 
\begin{deluxetable}{c|c|c|c|c|c|c|c|c|c}
\tablehead{
\colhead{parameter} & \dcolhead{\rho_{s}} & \dcolhead{p_{\rho}} &  \dcolhead{\tau_{k}} & \dcolhead{p_{\tau}} &
\colhead{parameter} & \dcolhead{\rho_{s}} & \dcolhead{p_{\rho}} &  \dcolhead{\tau_{k}} & \dcolhead{p_{\tau}}
}
\caption{Statistical Tests Results\label{tab:linearstats}}
\startdata
$g_{\rm p}$ & 0.44 & 0.03 & 0.31 & 0.03 & $R_{*}$ & -0.11 & 0.59 & -0.07 & 0.62 \\ [0.5 ex]
$H$ & 0.42 & 0.04 & 0.31 & 0.03 & $L_{\rm bo}$ & -0.11 & 0.60 & -0.07 & 0.62\\ [0.5 ex]
$\rho_{\rm p}$ & 0.41 & 0.04 & 0.30 & 0.04 & $M_{*} $ &  -0.11 & 0.62 & -0.08 & 0.57\\ [0.5 ex]
$\sqrt{T}/g$ & 0.41 & 0.04 & 0.28 & 0.05 &  $F_{\rm FUV} $ &  0.10 & 0.64 & 0.08 & 0.59\\ [0.5 ex]
$e$ & 0.37 & 0.07 & 0.28 & 0.06 & $T_{\rm eff}$ & -0.09 & 0.65 & -0.08 & 0.59\\ [0.5 ex]
$t_{\rm age}$ & 0.35 & 0.09 & 0.28 & 0.06 &$g_{*}$ & 0.09 & 0.66 & 0.08 & 0.56\\ [0.5 ex]
$M_{\rm p}$ & 0.21 & 0.31 & 0.13 & 0.37 & $T_{\rm eq}$ & 0.09 & 0.67 & 0.07 & 0.63\\ [0.5 ex]
$P_{\rm rot}$& 0.21 & 0.31 & 0.12 & 0.39 &$\rho_{*}$ & 0.08 & 0.71 & 0.06 & 0.69\\ [0.5 ex]
$f_{\rm HHe}$ & -0.16 & 0.44 & -0.11 & 0.44 & $F_{\rm XUV} $ & -0.06 & 0.77 & -0.05 & 0.73\\ [0.5 ex]
$R_{\rm p}$ &-0.16 & 0.45 & -0.10 & 0.50 &$Z_{\rm [Fe/H]}$ & -0.06 & 0.77 & -0.04 & 0.76\\ [0.5 ex]
$a$ &0.15 & 0.47 & 0.15 & 0.30 & $F_{\rm NUV}$ & 0.05 & 0.82 & 0.04 & 0.80\\ [0.5 ex]
\enddata
\tablecomments{We show the non-parametric Spearman's rho ($\rho_{s}$) and Kendall's tau ($\tau_{k}$) test results and the associated p-values ($p_{\rho}$ and $p_{\tau}$).}

\end{deluxetable}

\subsection{Stellar Forcing Parameters}

\begin{figure}
    \centering
    \includegraphics[scale = 0.4]{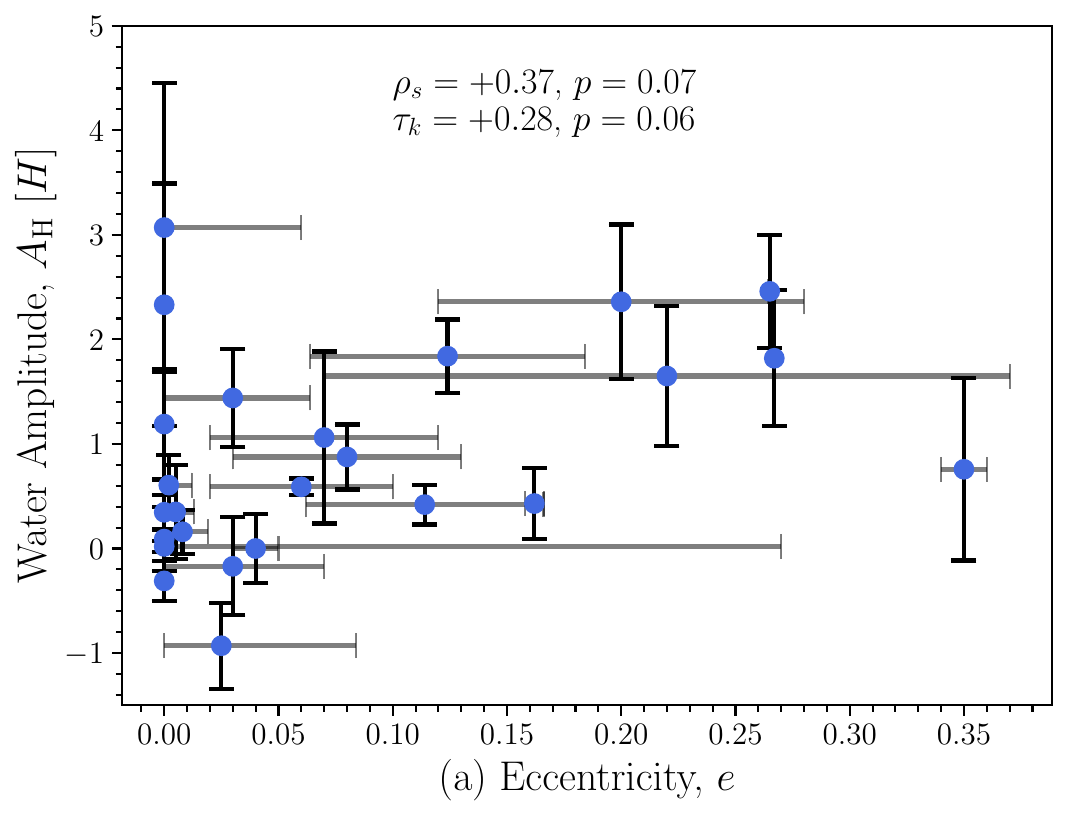}
    \includegraphics[scale = 0.4]{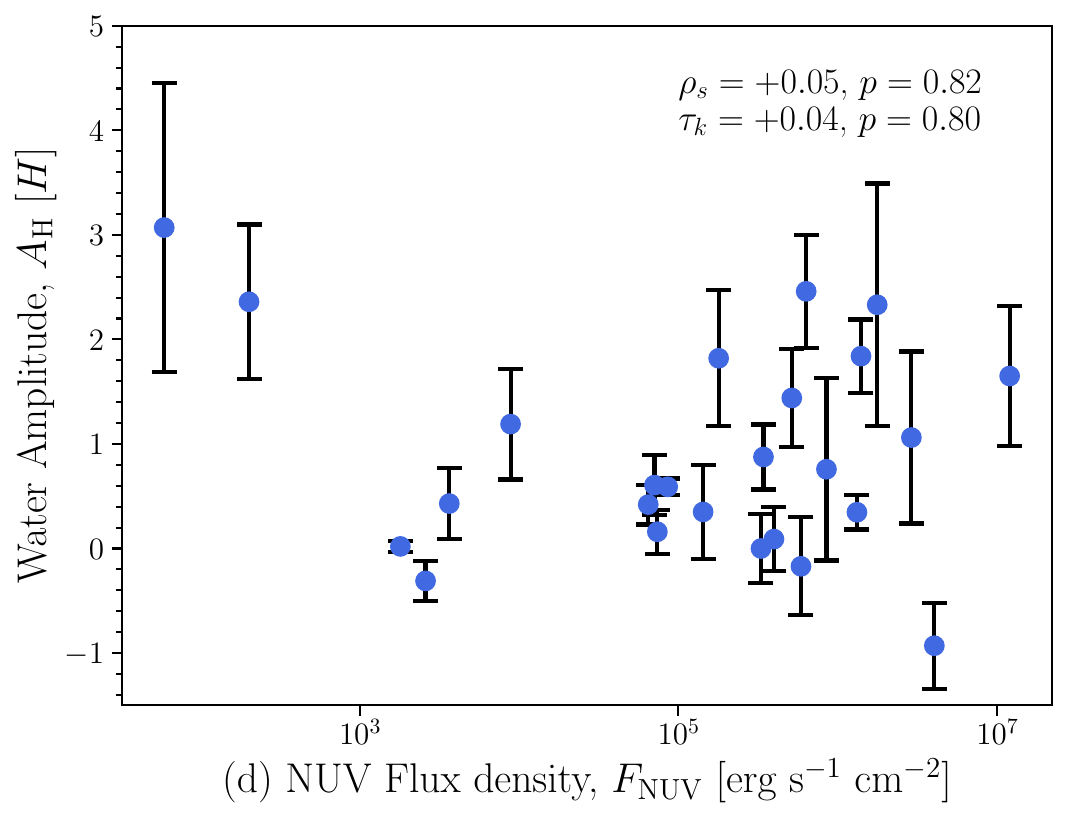}
    \includegraphics[scale = 0.4]{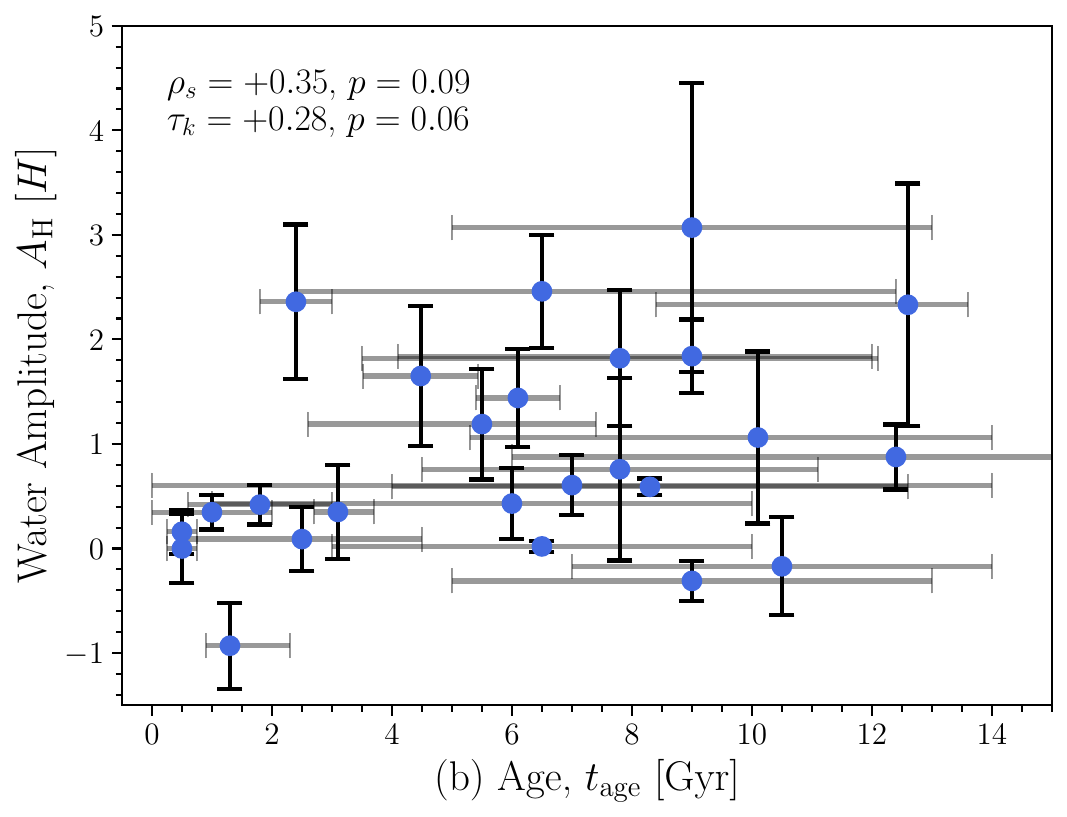}
    \includegraphics[scale = 0.4]{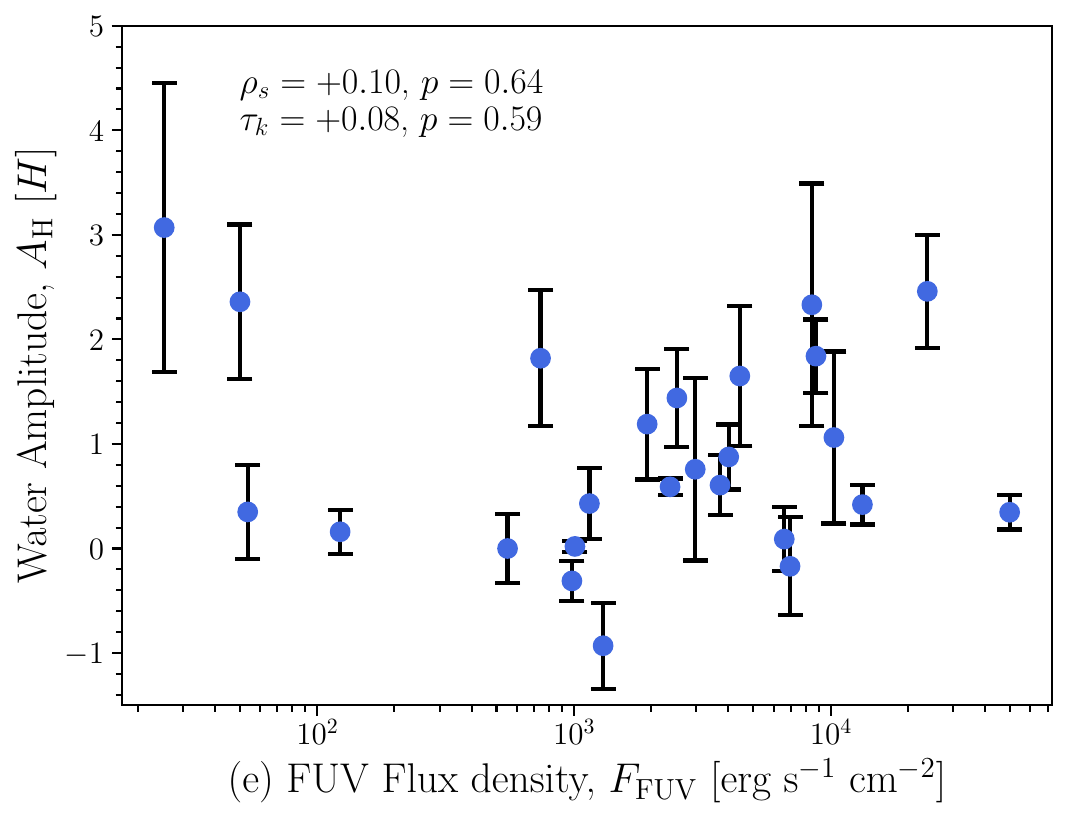}
    \includegraphics[scale = 0.4]{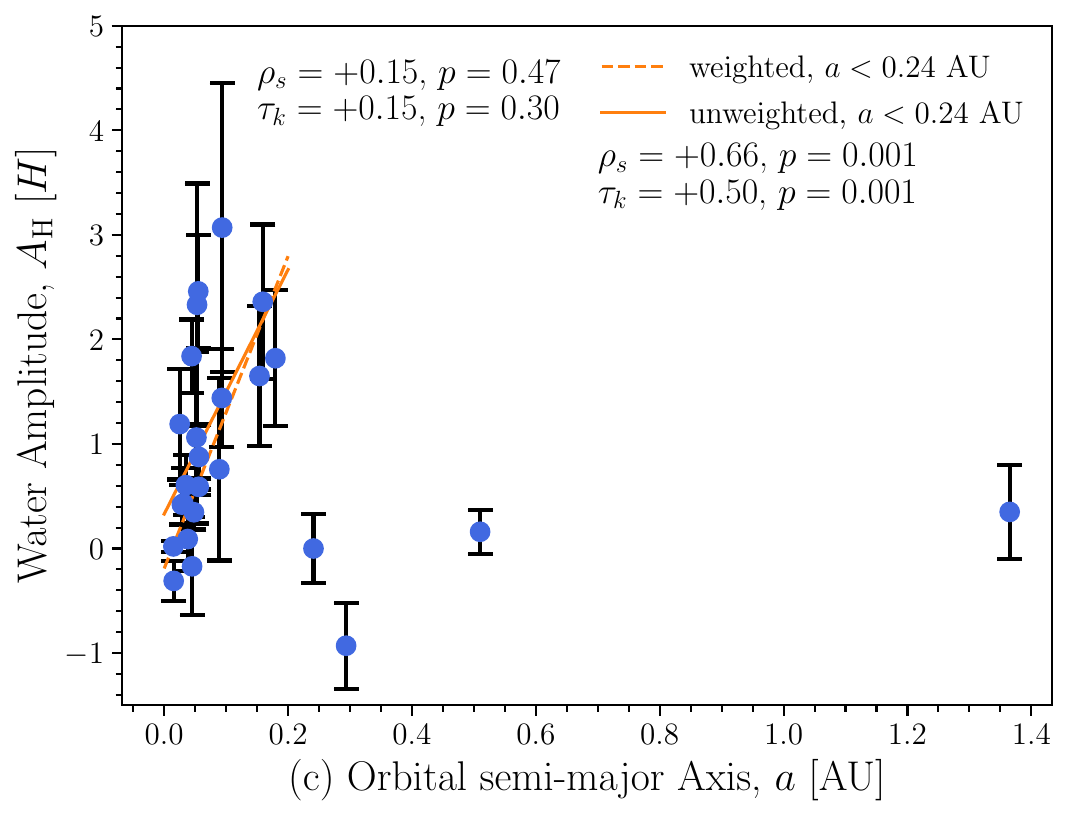}
    \includegraphics[scale = 0.4]{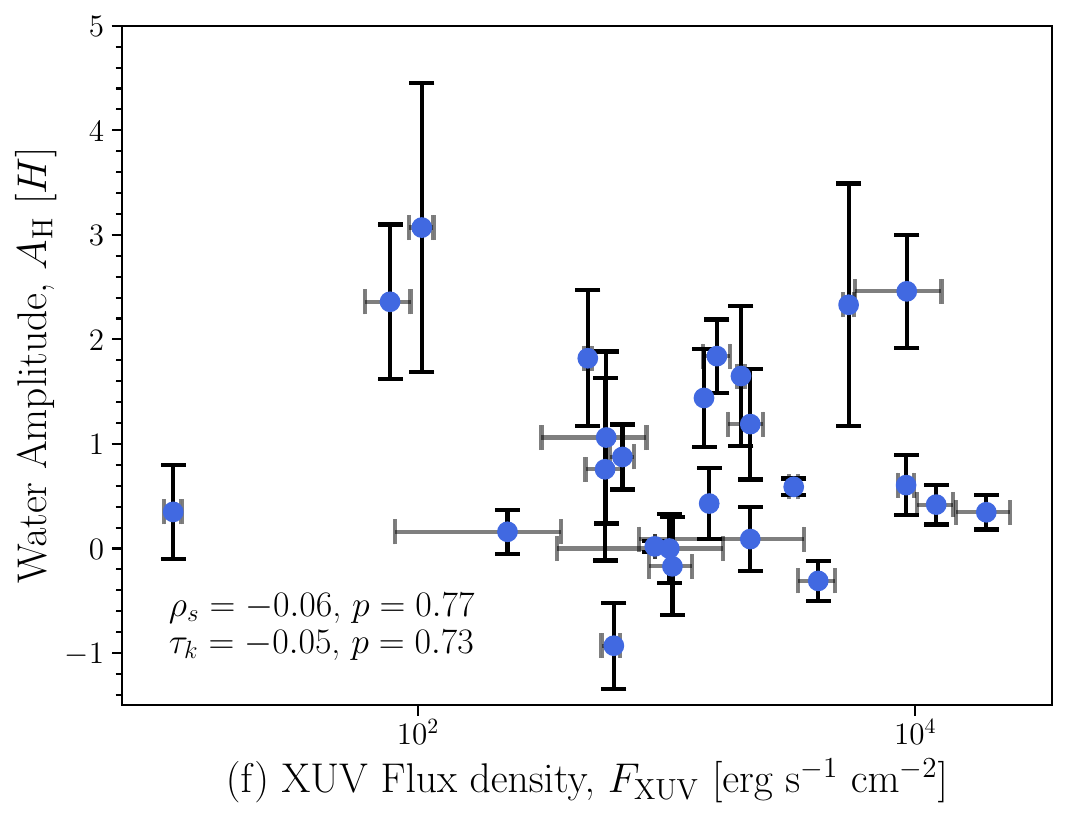}   
    \caption{\small (a)-(c): The tentative correlations among all the stellar forcing parameters, between $A_{\rm H}$ and orbital eccentricity ($e$), age of the star ($t_{\rm age}$) and orbital semi-major axis excluding the super-puff planets ($a$). The vertical error bars represent the estimated 1$\sigma$ uncertainty of $A_{\rm H}$, and the horizontal error bars (gray) represent 1$\sigma$ uncertainty of the parameter. The orange lines show the unweighted (solid) and weighted (dashed) linear fit using a least-squares fit, excluding Kepler 51b, d, Kepler 79, \& HIP 41378 f ($p<0.01$), with the associated non-parametric statistical tests below the line legend. (d)-(f): There is no correlations between $A_{\rm H}$ and the received stellar UV fluxes of the exoplanets.}
    \label{fig:stellar}
\end{figure}

Among all the stellar forcing parameters, orbital eccentricity ($e$) and age of the star ($t_{\rm age}$) hold tentative correlations against $A_{\rm H}$. The orbital semi-major axis ($a$) holds a relatively strong correlation against $A_{\rm H}$ ($p<0.01$), if we exclude the four low-density super-puff exoplanets. We do not see any noticeable correlations for all the other stellar forcing parameters, including mass ($M_*$), radius ($R_*$), density ($\rho_*$), surface gravity ($g_*$), effective temperature ($T_\text{eff*}$), bolometric luminosity ($L_{\rm bol}$), metallicity ($Z_{\rm [Fe/H]}$), stellar rotation period ($P_{\rm rot}$), and various stellar UV fluxes ($F_{\rm NUV}$, $F_{\rm FUV}$, $F_{\rm XUV}$), as shown in Table \ref{tab:linearstats}. In Figure \ref{fig:stellar}a-c, we show the test statistics between $A_{\rm H}$ and $e$, $t_{\rm age}$, and $a$. For $a$, both with and without the super-puff planets. In Figure \ref{fig:stellar}d-f, we show the relationships between $A_{\rm H}$ and the UV fluxes of different wavelengths. Plots for the rest of the stellar forcing parameters can be found in Figure \ref{fig:extra} in Appendix \ref{ap:B}.

Even though all the stellar forcing hold no statistically significant correlations with $A_{\rm H}$, one of the higher tentative correlations we found is between orbital eccentricity ($e$) and $A_{\rm H}$ (see Figure \ref{fig:stellar}a, $\rho_s = 0.37$, $p_{s} = 0.07$; $\tau_k = 0.28$, $p_k= 0.06$). This tentative correlation suggests exoplanets with circular orbits tend to be hazier than exoplanets with more elliptical orbits. This correlation is improved with the removal of the three planets with circular orbits and high water amplitude (i.e., TOI-674 b, LHS 1140 b, \& WASP-67 b). However, these planets hold no physical trait(s) in common to exclude them as population outliers. These planets host large uncertainties in $A_{\rm H}$ which indicate they could hold to the proposed trend. Additionally, the orbital eccentricity for many of the planets is highly unconstrained or assumed to be zero in the existing literature \citep[e.g.,][]{2021AA...653A..60M}. Better constrains on $e$ for many of these planets and more data points are needed to determine if this tentative correlation holds true.
In addition to orbital eccentricity, the other very tentative correlation we found is between stellar age ($t_{\rm age}$) and $A_{\rm H}$ (see Figure \ref{fig:stellar}b, $\rho_s = 0.35$, $p_{s} = 0.09$; $\tau_k = 0.28$, $p_k= 0.06$). Due to the sparseness of reliable age data from literature, the values we collected vary drastically in accuracy. In general, there appears to be a positive correlation where younger systems have hazier planets while older systems have clearer planets. But we also note that our sample did not holds no correlation between the $A_{\rm H}$ and the stellar metallicity (see Figure \ref{fig:extra} in Appendix \ref{ap:B}).

The orbital semi-major axis of the exoplanet ($a$) holds no correlation ($\rho_s = 0.15$, $p_{s} = 0.47$; $\tau_k = 0.15$, $p_k= 0.30$) with $A_{\rm H}$ (Figure \ref{fig:stellar}c). However, if we exclude the low-density super-puff targets (Kepler-51 b, d, Kepler-79 d, \& HIP 41378 f), we find a relatively strong correlation that is significant with $\tau_k$ and significant with $\rho_s$ ($\rho_s = 0.66$, $p_{s} = 0.001$; $\tau_k = 0.50$, $p_k= 0.001$). In Figure \ref{fig:stellar}c, we show this observed trend by plotting in orange the weighted and unweighted linear fit using a least-squares fit. This trend excluding the super-puffs indicates exoplanets closer to their host star are hazier than those further away. This agrees with physical intuition as closer exoplanets receive more relative stellar flux from this host star, which would help the haze formation processes. However, it is difficult to determine if the super-puff targets are outliers due to their extraordinary low bulk-densities \citep[][]{2016ApJ...817...90L} or if they actually belong to an overall nonlinear trend.

Stellar UV fluxes are believed to play a significant role in haze formation \citep[][]{Trainer18035} and are parameterized in models to scale for haze production \citep[][]{Kawashima&Ikoma18, 2019ApJ...878..118L}. However, we found no correlations between NUV, FUV, XUV fluxes and $A_{\rm H}$ ($p\ga0.60$, see also Figure \ref{fig:stellar}d-f). The $F_{NUV}$ and $F_{FUV}$ trends share a similar parabolic shape, as seen in Figure \ref{fig:stellar}d and \ref{fig:stellar}e. This shape is close to the trend found with $T_{\rm eq}$ (see Figure \ref{fig:planetary}b). This is not unexpected as we estimated the FUV/NUV fluxes of many of the targets' host stars (due to the lack of observed FUV/NUV data) using relationships that depend upon $T_{\text{eff}*}$ and $a$ (see Equation \eqref{eq:NUV}), which is also how $T_{\rm eq}$ is calculated. Specifically, $F_{NUV}$ and $F_{FUV}$ both have a positive correlation with $T_{\rm eq}$ ($\rho_{\rm s}=0.71$, $\tau_{\rm k} =0.55$ \& $\rho_{\rm s}=0.91$, $\tau_{\rm k} =0.85$, all $p$'s $ < 0.01$, respectively). For $F_{XUV}$, we find a no correlation (see Figure \ref{fig:stellar}f), in agreement with \citetalias{2017AJ....154..261C}. We still need more UV flux observations or better estimations for these exoplanet host stars (mostly M- type stars) to better assess these parameters.

\subsection{Planetary Parameters}\label{sec:correlation_planet}
\begin{figure}
    \centering
    \includegraphics[scale = 0.4]{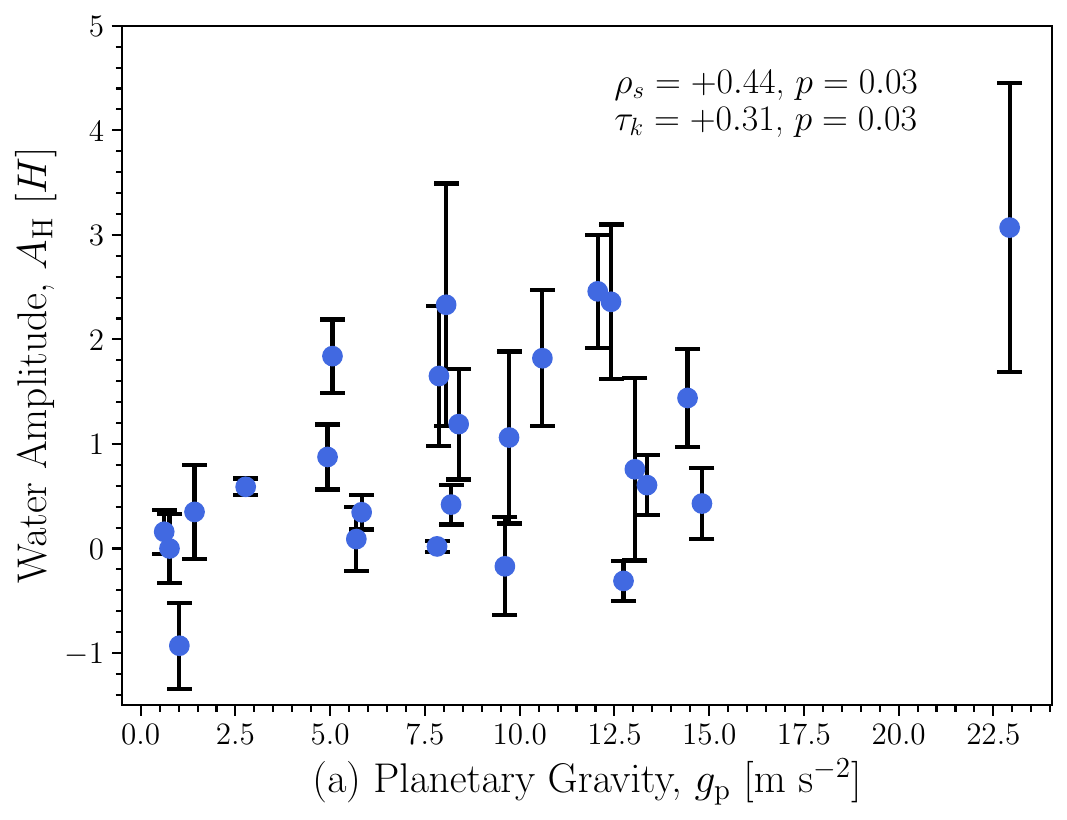} 
    \includegraphics[scale = 0.4]{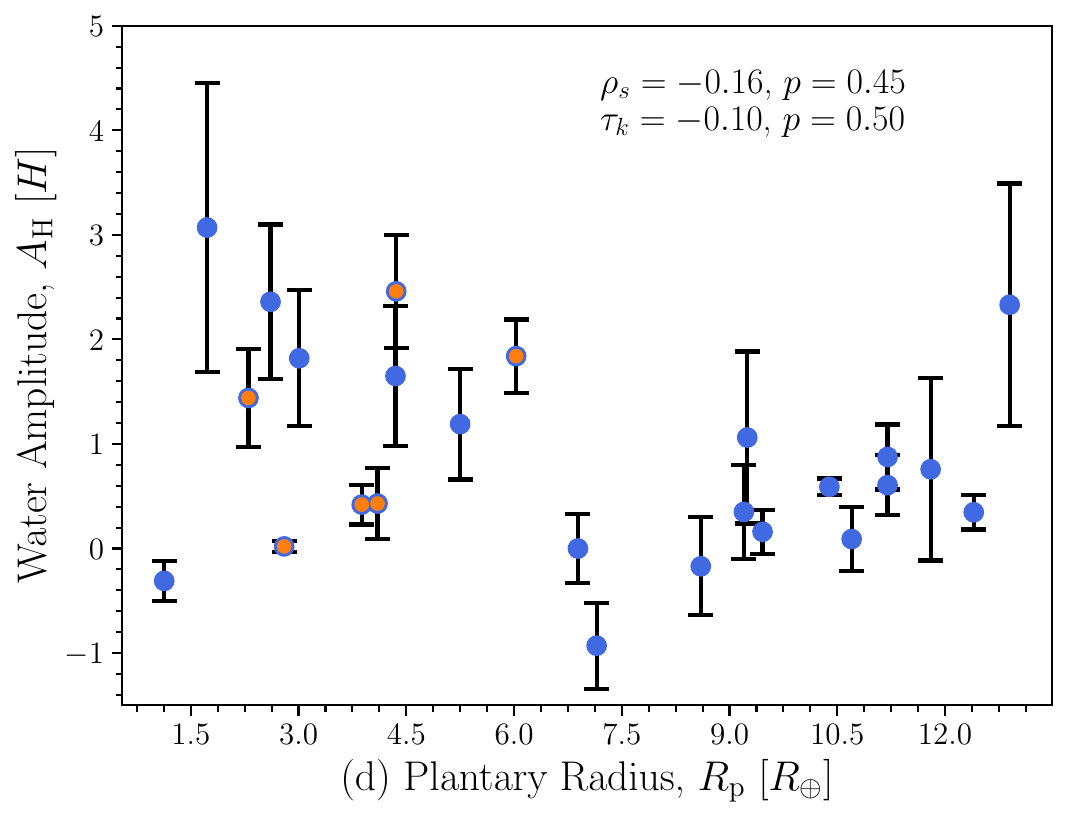}
    \includegraphics[scale = 0.4]{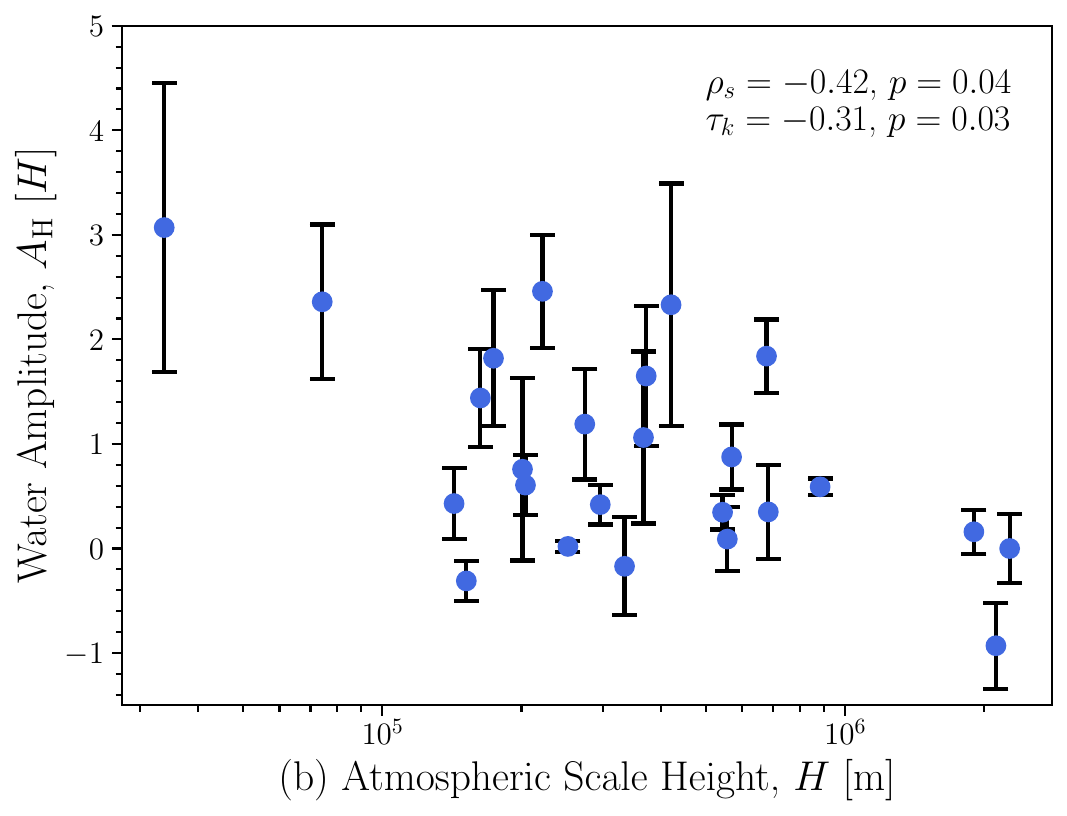}
    \includegraphics[scale = 0.4]{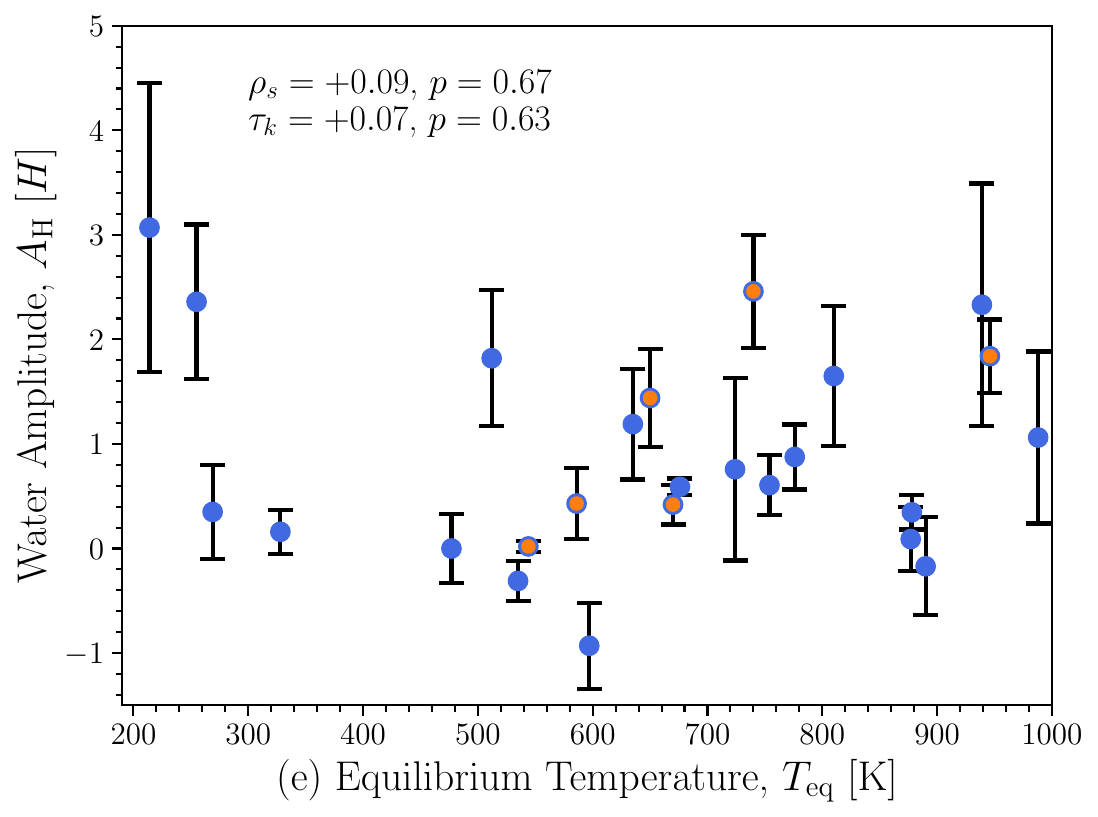}
    \includegraphics[scale = 0.4]{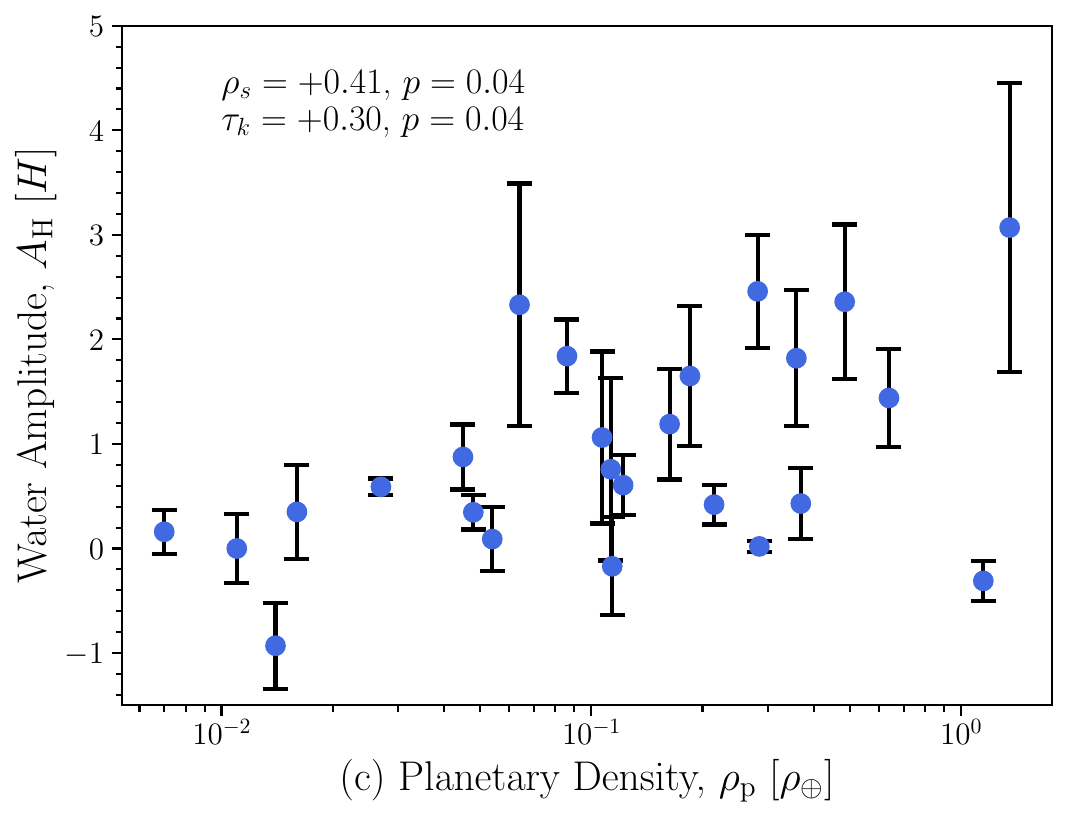}
    \includegraphics[scale = 0.4]{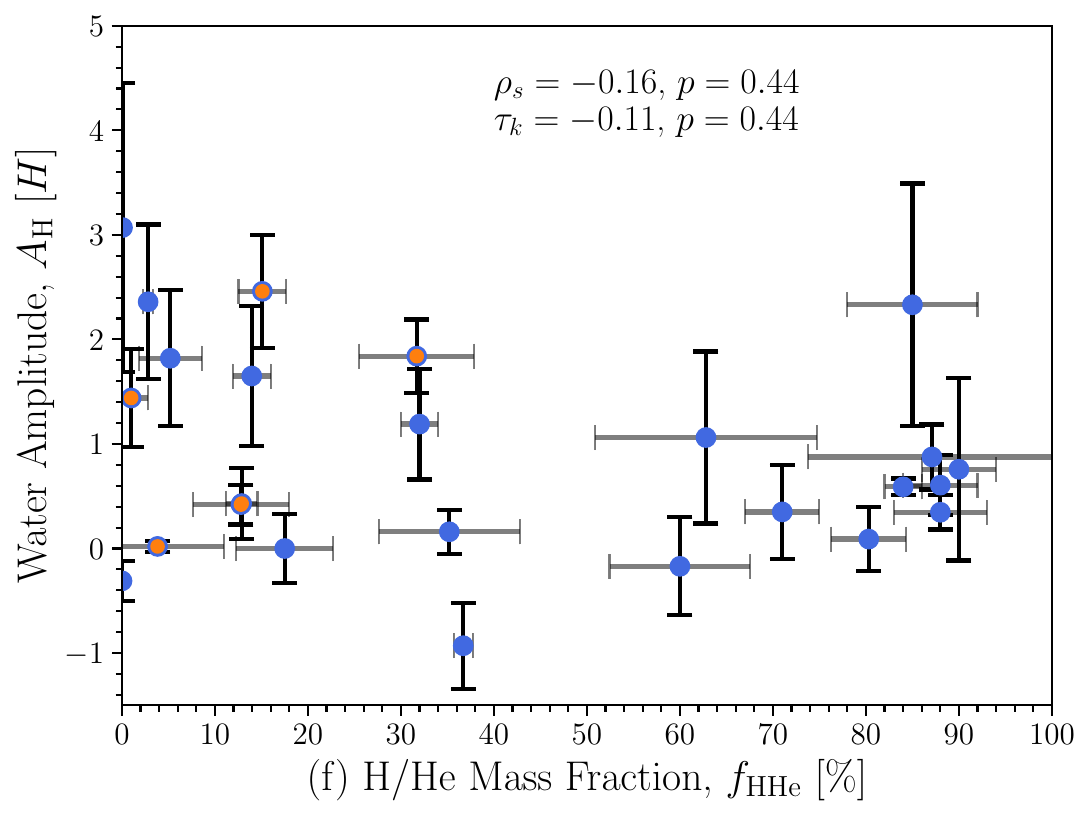}
    \caption{\small (a)-(c): Tentative correlations between $A_{\rm H}$ and planetary parameters, including atmospheric scale height ($H$), planet gravity ($g_{\rm p}$), and planet density ($\rho_{\rm p}$). The vertical error bars represent the estimated 1$\sigma$ uncertainty of $A_{\rm H}$, and the horizontal error bars (gray) represent 1$\sigma$ uncertainty of the parameter. (d)-(f): no correlations between $A_{\rm H}$ and planet radius ($R_{\rm p}$), planet equilibrium temperature ($T_{\rm eq}$), and planet atmospheric hydrogen-helium mass fraction ($f_{\rm HHe}$). The data points that are marked in orange are the targets in \citetalias{2017AJ....154..261C}. }
    \label{fig:planetary}
\end{figure}

Among all the planetary parameters, planet surface gravity ($g_{\rm p}$), atmospheric scale height ($H$), and planet density ($\rho_{\rm p}$) hold better but statistically insignificant correlations against $A_{\rm H}$ ($p\le0.04$). We do not see any noticeable monotonic correlations for all the other planetary parameters ($p>0.30$), including planetary mass ($M_{\rm p}$), radius ($R_{\rm p}$), equilibrium temperature ($T_{\rm eq}$), and bulk hydrogen-helium (H/He) mass fraction ($f_{\rm HHe}$), as summarized in Table \ref{tab:linearstats}. In Figure \ref{fig:planetary}a-c, we show the tentative correlations between $A_{\rm H}$ and $g_{\rm p}$, $H$, and $\rho_{\rm p}$. In Figure \ref{fig:planetary}d-f, we show the relationships for $R_{\rm p}$, $T_{\rm eq}$, and $f_{\rm HHe}$, which are found to be linear correlated with $A_{\rm H}$ in \citetalias{2017AJ....154..261C}. Plots for the rests of the planetary parameters can be found in Figures \ref{fig:extra} in Appendix \ref{ap:B}.

We find a tentative correlation ($\rho_s = 0.44$, $p_{s} = 0.03$; $\tau_k = 0.31$, $p_k= 0.03$) between $A_{\rm H}$ and planet gravity, $g_{\rm p}$ (see Figure \ref{fig:planetary}a). \citetalias{2016ApJ...817L..16S} found a weakly positive correlation for $A_{\rm H}$ with $g_{\rm p}$. However, \citetalias{2017ApJ...847L..22F} reported that they did not find any significant correlation between $A_{\rm H}$ and $g_{\rm p}$ with their larger hot-Jupiter data set. \citetalias{2017AJ....154..261C} also did not find significant correlation between $A_{\rm H}$ and $g_{\rm p}$ for the sub-Neptunes. This tentative correlation suggests that planets with higher gravity would have less hazy atmospheres, which could link to the removal of aerosol particles in planetary atmospheres. Larger gravity would lead to higher settling velocities of the aerosol particles, and could make the atmospheres clearer. We also identified a similar but slightly worse correlation between $A_{\rm H}$ and $\rho_p$ ($\rho_s = 0.41$, $p_{s} = 0.04$; $\tau_k = 0.30$, $p_k= 0.04$, see Figure \ref{fig:planetary}c).

We find a tentative correlation ($\rho_s = 0.42$, $p_{s} = 0.04$; $\tau_k = 0.31$, $p_k= 0.03$) between $A_{\rm H}$ and $H$ (see Figure \ref{fig:planetary}b). As demonstrated in Section \ref{sec:water_amplitude}, in our definition of the water amplitude metric, the transit depth difference between 1.25 $\mu$m and 1.4 $\mu$m was divided by scale height to remove the $H$ dependence (see Equation \eqref{eq:Ah} and \eqref{eq:Ah2}). However, $A_{\rm H}$ is still tentatively correlated with $H$. This suggests that an exoplanet atmosphere is likely hazier if its atmosphere is more extended. In Section \ref{sec:b_analysis}, we further explore whether $A_{\rm H}$ has any implicit dependence on $H$ using a simplified haze model. However, if we remove K2-18 b and LHS 1140 b and the super-puffs, which have the lowest and highest $H$ among all the targets, this trend becomes insignificant. We still need more atmospheric characterization data for exoplanets with intermediate $H$ (larger than the super-Earths and smaller than the super-puffs) to determine if this tentative trend holds true. 

The previously identified linear correlations no longer hold in our extended planet sample.
While \citetalias{2017AJ....154..261C} identified positive linear correlation between $A_{\rm H}$ and planet radius ($R_{\rm p}$), the addition of new planets breaks down this linear trend, as shown in Figure \ref{fig:planetary}d ($\rho_s = -0.16$, $p_{s} = 0.45$; $\tau_k = -0.10$, $p_k= 0.50$). 
\citetalias{2017AJ....154..261C} found a significant linear correlation ($r = 0.81$, $p = 0.05$; $\chi^2_{\nu} = 1.4$) between $A_{\rm H}$ and the estimated planet atmospheric hydrogen-helium mass fraction, $f_{\rm HHe}$,  whereas we find no significant linear correlation ($\rho_s = -0.16$, $p_{s} = 0.44$; $\tau_k = -0.11$, $p_k= 0.44$) between $A_{\rm H}$ and $f_{\rm HHe}$ with the addition of new planets (see Figure \ref{fig:planetary}f).
The linear correlation identified for $T_{\rm eq}$ in \citetalias{2017AJ....154..261C} also does not hold with the addition of new planets (Figure \ref{fig:planetary}e).

Recently, \citet{2021NatAs...5..822Y} pointed a possibly quadratic trend between $T_{\rm eq}$ and $A_{\rm H}$. They suggested that the exoplanets are likely haziest around 400--500 K because of the low removal rates of photochemical hazes. 
The low removal rate is due to a combination of the low surface energy of experimental haze analogs at $400$--$500~{\rm K}$ \citep{2021NatAs...5..822Y} and scarcity of condensable species in atmospheres with $T_{\rm eq}$ between 300--600 K \citep[][]{2020RAA....20...99Z}.
Meanwhile, hazes can be removed more easily for planets with higher and lower $T_{\rm eq}$, leading to clearer atmospheres. Thus, the previously identified positive linear correlation between $A_{\rm H}$ and $T_{\rm eq}$ in \citetalias{2017AJ....154..261C} could just be part of the quadratic trend for planets with $T_{\rm eq}>500$ K. \citetalias{2017ApJ...847L..22F} also found a positive correlation of $T_{\rm eq}$ with $A_{\rm H}$ for planets with $T_{\rm eq}$ between 500 and 2500 K. Since currently we only have two data points for the colder, clearer part of the quadratic trend (K2-18 b and LHS 1140 b), more transmission spectra data for colder planets are needed to solidify this proposed quadratic trend. 

\section{Analysis Motivated by Haze Microphysics}\label{sec:b_analysis}
In Section \ref{sec:water_amplitude}, we have demonstrated that $A_{\rm H}$ does not have any explicit linear dependence on any particular planetary/stellar forcing parameters. However, in Section \ref{sec:a_analysis}, we found that the water amplitude may be tentatively correlated with a few planetary/stellar forcing parameters. Interestingly, atmospheric scale height ($H$) has the strongest, although tentative, correlation with $A_{\rm H}$. From Equation \eqref{eq:Ah2} we can see that the $A_{\rm H}$ should not depend on the planetary scale height if the opacity is independent of the scale height.
This tentative correlation for $H$ potentially suggests that the haze opacity is dependent of the scale height.
In what follows, we further explore this possibility using a simple yet physically motivated haze model.

For a clear atmosphere with water vapor, $A_{\rm H}\approx 7$ \citepalias{2017AJ....154..261C} due to the molecular opacity difference of water at 1.25 $\mu$m and 1.4 $\mu$m. In Section \ref{sec:a_analysis}, we found none of our exoplanet targets have $A_{\rm H}$ values close to 7, indicating the ubiquity of clouds and hazes in their atmospheres. 
Photochemical hazes are more likely the predominant opacity sources according to previous modeling studies \citep{2020NatAs...4..951G}.
Thus, we here examine what planetary/stellar parameter dependence the haze absorption causes in the water amplitude.
Co-dependency on haze absorption and molecular water absorption makes a general analytical solution for $A_{\rm H}$ like Equation \eqref{eq:Ah2} impossible. 
For simplicity, we assume the water absorption dominates the opacity at 1.4 $\mu$m and the haze absorption dominate the opacity at 1.25 $\mu$m.
Then, Equation \eqref{eq:Ah2} can be written as:
\begin{equation}
     A_{\rm H}=\ln{\left(\frac{\kappa_{\rm H_2O}(1.4~{\rm \mu m})}{\kappa_{\rm haze}(1.25~{\rm \mu m})}\right)},
\label{eq:Ah3}
\end{equation}
where the opacity of molecular water ($\kappa_{\rm H_2O}$) is a constant at 1.4 $\mu$m, and the haze opacity ($\kappa_{\rm haze}$) can be analytically approximated as \citep{Ohno&Kawashima20}:
\begin{equation}\label{eq:kappa_haze1}
    \kappa_{\rm haze}=\frac{36\pi g_pHF_{\rm haze}}{\rho_{\rm haze}Pv_{\rm t}}\left[ 1-\exp{\left(-\frac{v_{\rm t}H}{K_{\rm zz}} \right)}\right]\frac{nk\lambda^{-1}}{(n^2-k^2+2)^2+(2nk)^2},
\end{equation}
where $F_{\rm haze}$ is the downward haze mass flux from its chemical source in the upper atmosphere, which corresponds to the column-integrated haze production rate, $K_{\rm zz}$ is the eddy diffusion coefficient, $P$ is the atmospheric pressure, $\rho_{\rm haze}$ is the density of the haze particles, $n$ and $k$ are wavelength-dependent real and imaginary parts of the refractive indices of the hazes, and $v_{\rm t}$ is the terminal velocity of haze particles. For tiny haze particles, we can approximate $v_{\rm t}$ as \citep[][]{2003A&A...399..297W}:
\begin{equation}\label{eq:vt}
v_{\rm t} \approx \frac{\rho_{\rm haze}g_{\rm p}^2H}{P\sqrt{8k_BT/\pi{\mu}}}a_{\rm haze},
\end{equation}
where $a_{\rm haze}$ is the radius of haze particles. Equation \eqref{eq:kappa_haze1} is valid when the haze particles have low single scattering albedos and are much smaller than the gas mean free path and the relevant wavelengths. Additionally, $K_{zz}$ and $a_{\rm haze}\rho_{\rm haze}$ were assumed to be constant when deriving Equation \eqref{eq:kappa_haze1} \citep{Ohno&Kawashima20}.

Since the eddy diffusion coefficient is still highly uncertain for exoplanets, here we assume that the eddy diffusion timescale is much slower than the gravitational settling timescales, i.e., $v_tH/K_{zz} \gg 1$. 
By assuming constant haze density, particle size, planetary gravity, and temperature, we can now further reduce the haze opacity of Equation \eqref{eq:Ah3} into:
\begin{equation}\label{eq:kappa_haze2P}
    \kappa_{\rm haze} = 
           \frac{36\sqrt{8\pi k_{\rm b}}}{\sqrt{\mu}} \frac{F_{\rm top}f\lambda^{-1}}{a_{\rm haze}\rho_{\rm haze}^2}\frac{\sqrt{T}}{g_{\rm p}},   \quad  v_tH/K_{zz} \gg 1,\\
\end{equation}
where $f=nk/[(n^2-k^2+2)^2+(2nk)^2]$ is a dimensionless constant depending on haze refractive indices.
Plugging Equation \eqref{eq:kappa_haze2P} into Equation \eqref{eq:Ah3}, $A_{\rm H}$ can be expressed as:
\begin{equation}\label{eq:kappa_haze2}
    A_{\rm H} = -\ln
       \left(\frac{36\sqrt{8\pi k_{\rm b}}}{\sqrt{\mu}} \frac{F_{\rm top }f\lambda_{\rm 1.25}^{-1}}{a_{\rm haze}\rho_{\rm haze}^2\kappa_{\rm H2O}}\frac{\sqrt{T}}{g_{\rm p}}\right),
\end{equation}
where $\lambda_{\rm 1.25}=1.25~{\rm {\mu}m}$.
Equation \eqref{eq:kappa_haze2} indicates that the water amplitude becomes smaller as the temperature increases and/or the gravity decreases.
This is because high $T$ and low $g$ lead to a large atmospheric scale height, which hinders the removal of haze particles through eddy diffusion and gravitational settling.
Because the atmospheric scale height is proportional to $T/g$, Equation \eqref{eq:kappa_haze2} can be rewritten to show an explicit dependency with scale height and gravity, which partially agrees with our best tentative correlations found for the atmospheric scale height and gravity (Section \ref{sec:correlation_planet}).

\begin{figure}
    \centering
    \includegraphics[scale = 0.6]{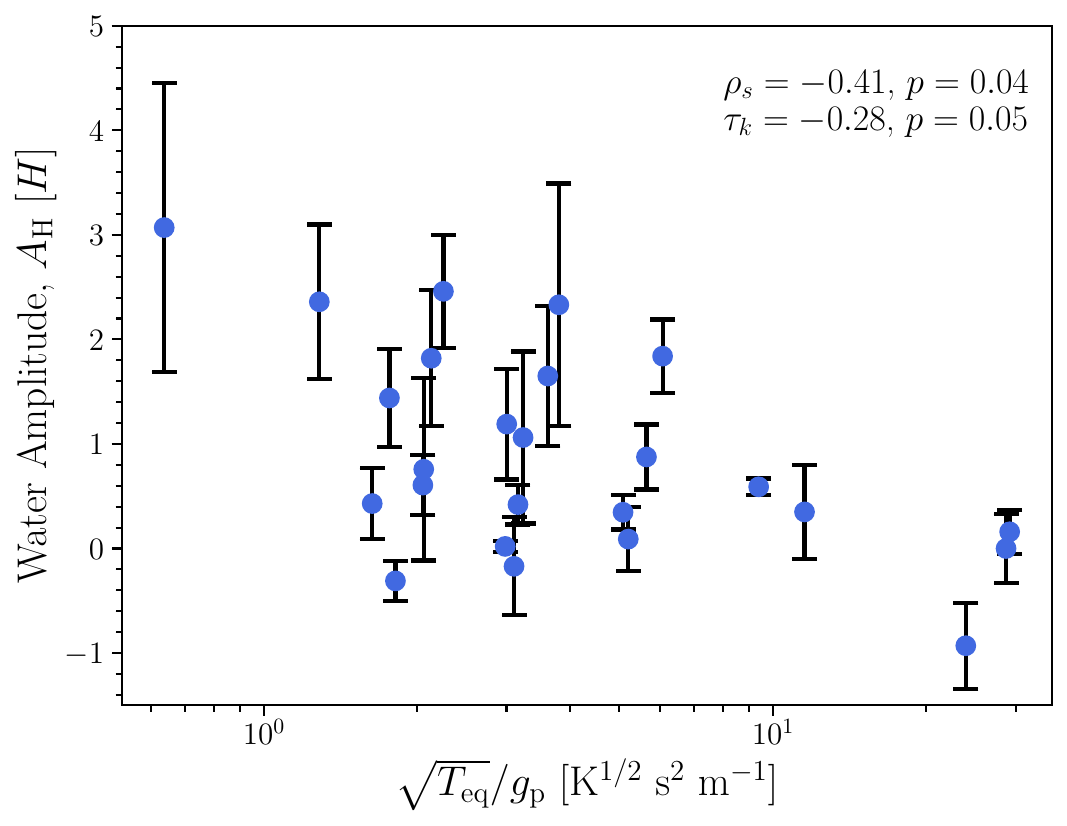}
    \caption{\small Correlations between $A_{\rm H}$ and $\sqrt{T_{\rm eq}}/g_p$. The vertical error bars (black) represent the estimated 1$\sigma$ uncertainty of $A_{\rm H}$.}
    \label{fig:T_g}
\end{figure}

Motivated by haze microphysics, we can use this theoretical metric using combinations of a few planetary parameters and examined its correlation with $A_{\rm H}$. Equation \eqref{eq:kappa_haze2} shows that $A_{\rm H}$ may have an implicit dependence on both planet gravity and temperature. 
Our previous assumptions for Equation \eqref{eq:Ah2} allow us to set $T$ as $T_{\rm eq}$ and $g_p$ as our vertically constant planet gravity. 
If $F_{\rm top}$ and $a_{\rm haze}$ are insensitive to planetary parameters, the water amplitude would have implicit dependence $\sqrt{T_{\rm eq}}/g_p$.
Thus, we further explored whether $A_{\rm H}$ has any correlation with this asymptotic solution ($\sqrt{T_{\rm eq}}/g_p$, see Figure \ref{fig:T_g}). A strong correlation $\sqrt{T_{\rm eq}}/g_p$ would indicate a preference for the settling regime for our planet set. 

For $\sqrt{T_{\rm eq}}/g_p$, we find a tentative correlation with $A_{\rm H}$ ($\rho_s = 0.41$, $p_{s} = 0.04$; $\tau_k = 0.28$, $p_k= 0.05$). This tentative correlation is one of the highest we find compared to the blind search for most of the stellar forcing and planetary parameters. However, none of these parameters can be considered to have statistically significant correlations with $A_{\rm H}$. The tentative correlation of $\sqrt{T_{\rm eq}}/g_p$ could be caused by several reasons listed below: 1) our exoplanet targets do not fall into the settling regime only; 2) we have fixed a number of uncertain parameters, such as downward haze mass flux $F_{\rm top}$, but they may also have dependence on planetary parameters; 3) we assume the atmospheric opacity at 1.4 $\mu$m is dominated by water and at 1.25 $\mu$m is dominated by hazes, while in reality, the opacity at both wavelengths should be the combination of both haze and gas opacity; 4) the calculated $A_{\rm H}$ for the exoplanet targets may have some intrinsic spread due to assumption we make in the calculations, such as assuming a constant mean molecular weight of the atmosphere ($\mu=2.3$ amu); 5) condensation clouds may also contribute to the atmospheric opacity in addition to photochemical haze for our exoplanet targets \citep[e.g.,][]{2013ApJ...775...33M}. 

Overall, our analysis led to an asymptotic solution that also is not significantly correlated with the derived $A_{\rm H}$, for our exoplanet targets. This supports that $A_{\rm H}$, or haziness of exoplanets, may not be solely determined by a single or a simple combination of planetary/stellar forcing parameters. 
We note that our haze model presented here is highly simplified. 
Further studies with detailed microphysical models \citep[e.g.,][]{Lavvas&Koskinen17,Kawashima&Ikoma18,2020ApJ...890...93G,Ohno&Tanaka21} will be warranted to examine the haze trends that may appear in observed correlations. 

\section{Discussion}\label{sec:discussion}

\subsection{Consistency with Previous Studies} 
Even though we have adopted a simpler method than \citetalias{2017AJ....154..261C} to compute the water amplitude, we confirmed that most of our $A_{\rm H}$ values are in close agreement with those derived by \citetalias{2017AJ....154..261C} if we use their planetary parameters.  We also confirmed that our $A_{\rm H}$ values are in agreement with those derived by \citetalias{2016ApJ...817L..16S, 2017ApJ...847L..22F} using the planetary values we collected. The only exception is HD 97658 b, which we calculated its $A_{\rm H}$ to be $1.44\pm0.50$, using the updated transmission spectra data from \citet{2020AJ....159..239G} (4 transits). While \citetalias{2017AJ....154..261C} gives $A_{\rm H}=-0.09\pm0.55$ using data from \citet{2014ApJ...794..155K} (2 transits).

\subsection{Caveats}
We have adopted several assumptions in our analysis.
In the following sections, we discuss the validity of our assumptions along with potential caveats for our sample and data analysis.

\subsubsection{Statistical Uncertainties}
In Section \ref{sec:a_analysis}, we do not consider the effects of our relatively small sample size on the test statistics. Even though we increased our exoplanet sample size from 6 to 25 compared to \citetalias{2017AJ....154..261C}, our sample size is still tiny compared to the confirmed planet population size ($\sim$0.5\% of the over 5000 confirmed exoplanets).  Our small sample size and relative errors on $A_{\rm H}$ and many parameters makes outliers difficult to catch and more difficult to statistically rule out. For example, our super-puff targets present the most outlier-like behavior and are physically the most different (see Figure \ref{fig:stellar}c). However, it is difficult to conclude whether they represent end-member behaviors of a continuous trend or whether they are outliers. In Section \ref{sec:sst}, we discuss the inclusion of the super-earth targets (LHS 1140 b \& GJ 1132 b) and their effects on the trends. 

For these reasons, we used the non-parametric statistical tests, Spearman's rho and Kendall's tau, in order to accurately determine any level of correlation between our parameters and $A_{\rm H}$. These tests point to none of our parameters hold any statistically significant correlations with $A_{\rm H}$. Thus, the tentative trends we discussed in Section \ref{sec:a_analysis} are mere indicators of potential correlations between stellar forcing/planetary parameters and exoplanet haziness to aid future observation and modeling efforts. More observations from the \textit{HST} and future missions such as the \textit{James Webb Space Telescope} (\textit{JWST}) and the \textit{Atmospheric Remote-sensing Infrared Exoplanet Large-survey} (\textit{ARIEL}) will alleviate some of this uncertainty by increasing the number of exoplanets measured through transmission spectroscopy. 
 
\subsubsection{Selection of Atmospheric Metallicity and Albedo}\label{sec:albedo}
When calculating $A_{\rm H}$ (see Equation \eqref{eq:Ah}), we divide the transit radii difference at 1.4 $\mu$m and 1.25 $\mu$m by the atmospheric scale height to remove the explicit planetary parameter dependence. However, when determining the atmospheric scale height for each of our exoplanet targets, we make two key assumptions: 1) all exoplanets have a H$_2$-He dominated atmospheres with atmospheric mean molecular weight $\mu=2.3$ amu, and 2) they all have a planetary Bond albedo $A_B=0.3$. While some sub-Neptunes are suggested to have atmospheric metallicity close to solar values \citep[e.g., GJ 3470 b, ][]{2019NatAs...3..813B}, some sub-Neptunes likely have much higher metallicity ($\ga100$ times solar metallicity), such as GJ 1214 b \citep{2013ApJ...775...33M, Morley+15, Ohno&Okuzumi18, Gao&Benneke18, 2019ApJ...878..118L, Ohno+20a} and GJ 436 b \citep{2017AJ....153...86M}. 
Planet formation models also predicted that sub-Neptunes tend to have high-metallicity atmospheres \citep{2013ApJ...775...80F,Venturini+16,Cridland+19}.
Thus the first assumption may not hold true for all exoplanets. Previous trend studies have used $\mu$ values from 2.3 amu for all planets \citep{2016Natur.529...59S, 2017AJ....154..261C}, or 2.3 amu for Jupiter-sized exoplanets and 3.8 amu for Neptune-sized exoplanets and below \citepalias{2016ApJ...817L..16S, 2017ApJ...847L..22F}. Similarly, there is no typical expected values of $A_B$ for most of our exoplanet targets. Previous trend studies have used $A_B=0$ \citepalias{2016ApJ...817L..16S, 2017ApJ...847L..22F} or $A_B=0.2$ \citepalias{2017AJ....154..261C}.

Since $H$ is used for normalization, a change in $\mu$ or $A_B$ acts as a scale factor to increase or decrease $A_{\rm H}$. The errors of these systematic effects can be determined. For $\mu$, the percent error goes as $1-\mu_{2.3}/\mu_{\rm True}$. That gives for $\mu=3.8$ amu and $\mu=12$ amu a $\sim40\%$ and $\sim80\%$ error, respectively. Higher $\mu$ for some exoplanets may start to affect our analysis, and would mainly affect exoplanets with larger water amplitudes ($A_{\rm H} \ga 1.5$). For $A_B$, the percent error goes as $1-(1-A_{\rm B, True})^{1/4}/(0.915)$. That gives for $A_B=0$ a $\sim9.3\%$ error against $A_B=0.3$. Thus, the choice of $A_B$ has little affect on our analysis unless these exoplanets are highly reflective ($A_B\sim1$).

We attempted to remedy our assumption of $\mu$ by using an empirical mass-metallicity relationship of our solar-system giant planets \citep[][]{2017Sci...356..628W} and a metallicity-mean molecular weight relationship \citep[][]{2011ApJ...733....2N}. We can then assign each of our targets a unique mean molecular weight based on their planet mass. We then attempted to repeat the statistical analysis. For the parameters that held tentative correlations with A$_{\rm H}$ (atmospheric scale height, planetary surface gravity, and planet density, eccentricity), the calculated test statistics values changed minimally. For some of the better tentatively correlated parameters (e.g., scale height and planet density), the associated p-values decrease slightly, but none reach our significance threshold. Thus, our choice of $\mu = 2.3$ amu has minimal effect on our analysis. Although, better constraints on $\mu$ may become more important when more terrestrial-like exoplanets are added to our sample.
 
\subsubsection{Terrestrial Atmospheres}\label{sec:sst}
\begin{figure}
    \centering
    \includegraphics[scale=0.49]{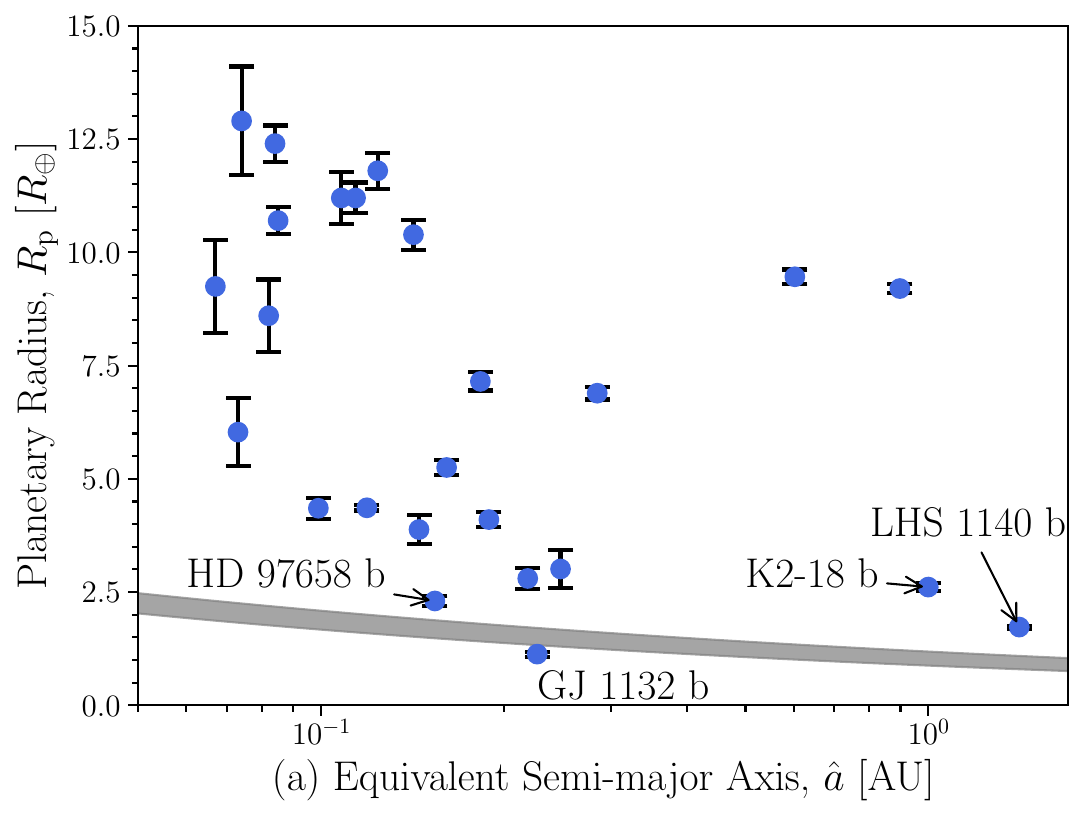}
    \includegraphics[scale=0.49]{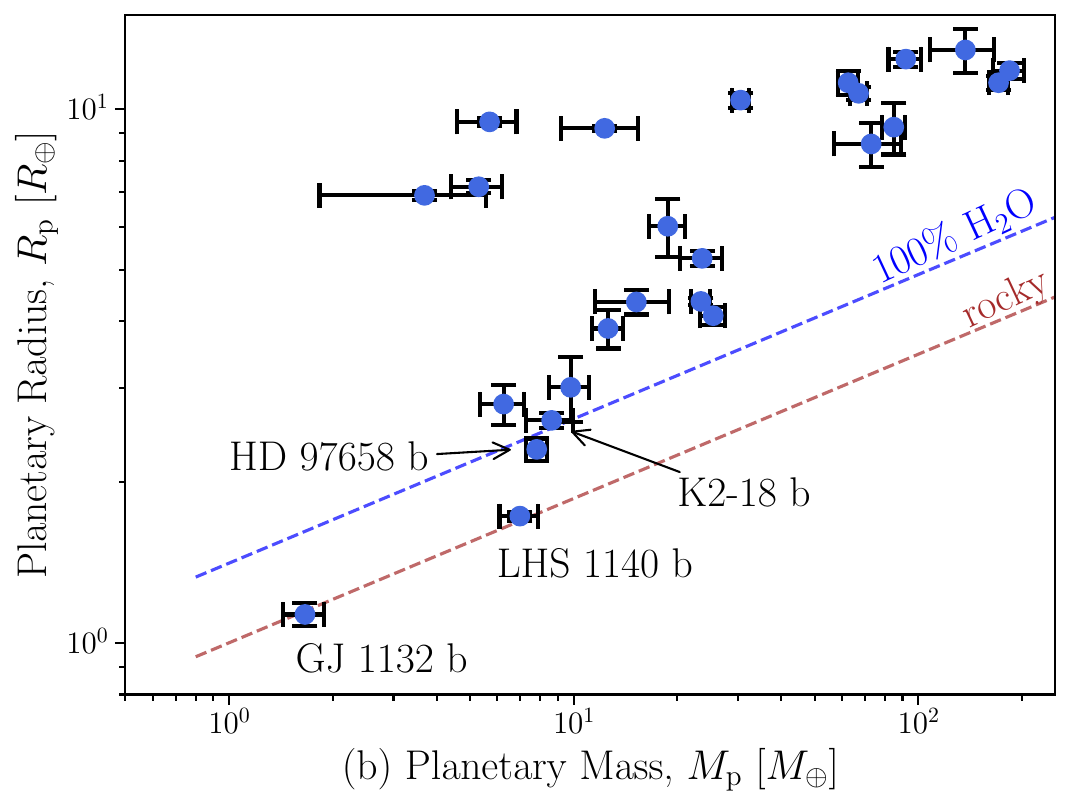}
    \caption{\small 
    (a) Determining whether our exoplanets targets are within the radius valley. The shaded region represents a fit to the radius valley where the atmosphere is completely lost due to evaporation \citep[adapted from ][]{2020A&A...638A..52M}. The orbital distances of the targets are scaled to an equivalent semi-major axis, assuming they are orbiting the same Sun-like star ($T_{\rm eff*} = 5778$ K and $R_* = R_{\rm sun}$).
    (b) Mass-Radius distribution of our exoplanets targets compared to 100\% H$_2$O (blue dashed line) atmosphere and Earth-like rocky (brown dashed line). Both lines are calculated from $R_{\rm p} [R_{\oplus}] = (1 +0.55x - 0.14x^2)(M_{\rm p} [M_{\oplus}])^{1/3.7}$, where $x$ is the ice mass fraction ($x=1$ for 100\% H$_2$O, $x=0$ for rocky) \citep[adapted from][]{Zeng9723}.
    }
    \label{fig:ahat}
\end{figure}

While we expect most of our targets to hold H$_2$-He-rich atmospheres, some of our smaller targets could have atmospheres that are more terrestrial-like (e.g., outgassed secondary atmospheres or super-high metallicity primordial atmospheres). We first examined whether our targets are below or above the proposed radius valley by using the relationship proposed by \citet{2020A&A...638A..52M} (see Fig \ref{fig:ahat}a). For planets below the radius valley, they are proposed to be close to stripped rocky cores due to the high stellar irradiation they receive \citep[][]{Lopez&Fortney13, 2013ApJ...775..105O, Owen&Wu17}
and are more likely to have thin terrestrial-like atmospheres. As shown in Figure \ref{fig:ahat}a, only GJ 1132 b is below the radius valley.
Thus, GJ 1132 b exists in a parameter space in which most of its atmosphere tends to be lost by photoevaporation.

We also examined the bulk densities of our targets to verify our assumption of H$_2$-He atmosphere.
As shown in Figure \ref{fig:ahat}b, most of our sample planets have radii fairly larger than those expected for pure H$_2$O planets, implying the presence of H$_2$-He atmospheres that puff up the observable radius.
However, the bulk densities of HD 97658 b and K2-18 b are consistent with those of 100\% H$_2$O planets. Thus, it is unclear whether these planets are rocky planets with H$_2$-He atmospheres or icy planets with steam-like (i.e., high metallicity) atmospheres.
The impacts of the possible high-metallicity atmospheres are discussed in Section \ref{sec:albedo}.
In addition, both LHS 1140 b and GJ 1132 b have Earth-like bulk densities. This presumably indicates that there are little-to-no H$_2$-He atmospheres in LHS 1140 b and GJ 1132 b. 
These targets may contribute two effects in our analysis: 1) terrestrial atmospheres likely have much higher $\mu$ than H$_2$-He atmospheres and their true atmospheric water amplitudes could greatly deviate from the ones calculated with $\mu=2.3$ amu (see discussion in Section \ref{sec:albedo}), 2) these terrestrial exoplanets could be the outliers of our samples, since their atmospheric chemistry could be very different from sub-Neptunes due to the differences in bulk atmospheric composition. Thus, we removed these two targets from our samples and reanalysed all the trends in Section \ref{sec:a_analysis}. Again, we found that our results changed minimally for most stellar forcing/planetary parameters. For some of the tentatively correlated parameters (e.g., eccentricity), the associated p-values decrease slightly, but none reach our significance threshold. Thus, we leave the both of the planets in our analysis.

\subsubsection{Spectral Signatures: Water vs Methane}\label{sec:spectral}
We note that while we call our metric the ``water" amplitude, any other molecules with spectral features present at around 1.4 $\mu$m may also be attributed to increase the ``water" amplitude. For example, \citet{2020arXiv201110424B} suggests that methane absorption (which also has a spectral feature of 1.4 $\mu$m) could dominate over water absorption for exoplanets with $T_{\rm eq}<600$ K. In this study, the ``water" amplitude is a mere measurement of the atmospheric haziness of exoplanets. The main absorber species at 1.4 $\mu$m does not affect our calculation $A_{\rm H}$ or assessment of the haziness of the exoplanet. Unless if the main absorber is substantially changing the mean molecular weight of the atmosphere, see the discussion in Section \ref{sec:albedo}.

\subsubsection{Transmission Data Reduction}
In this study, we collected transit data from the literature in which the data reduction methods vary for each study. Such inhomogeneous data reduction may affect the results of our analysis. For example, the choice of the number of channels averaged in each wavelength bin can disrupt small features and the differences of intermediate data reduction steps could completely change the resulting data product \citep[e.g.,][]{2018A&A...620A.142A}.
As an example, the transit data of GJ 1132 b was reduced independently by three groups:  \citet{2021AJ....161..213S,2021AJ....161...18M,2021arXiv210510487L}. \citet{2021AJ....161..213S} analysis produced a sloped spectral feature around 1.3 $\mu$m, and a larger negative $A_{\rm H} \approx -0.9$. The analysis of \citet{2021AJ....161..284M} and \citet{2021arXiv210510487L} found instead a visibly flat spectra, and with $A_{\rm H}$ close to zero. Since the last two groups return similar results, we choose to calculate $A_{\rm H}$ using reduced data from \citet{2021AJ....161..284M}, which is also close to the $A_{\rm H}$ calculated from \citet{2021arXiv210510487L}. A trend study using uniform data reduction methods would be preferred, allowing for more rigorous comparison between exoplanets intrinsic properties \citep[e.g.,][]{2016Natur.529...59S, 2018AJ....155..156T,2021AJ....162...37R}. At present, we choose reduced data from the more recently published works, which either incorporate data from more recent transit observations (e.g., HD 97658 b we use \citealp{2020AJ....159..239G} instead of \citealp{2014Natur.505...66K}) or use more up-to-date stellar parameters (e.g., K2-18b we use \citealp{2019ApJ...887L..14B} instead of \citealp{2019NatAs...3.1086T}). Future studies with a consistent data reduction pipeline for all targets would provide more rigorous correlations between $A_{\rm H}$ and planetary/stellar parameters.

\subsubsection{Negative Water Amplitudes}
For three of our targets, we find a negative water amplitude: GJ 1132 b, WASP-29 b, and Kepler 79 d. For GJ 1132 b and WASP-29 b, their $A_{\rm H}$ are close to zero ($A_{\rm H}=-0.31\pm0.19$ and $-0.17\pm0.47$, respectively) and in agreement with a flat spectra. Kepler 79 d is our only negative $A_{\rm H}$ of concern ($A_{\rm H}=-0.93\pm0.41$). This could be partly due to the uncertainty of transit depth from Kepler 79 d's limited number of transits, which could be improved with more transit observations.
For TRAPPIST-1 planets, the observed negative $A_{\rm H}$ are suggested to originate from the heterogeneity on the photosphere of their host star \citep{2018ApJ...853..122R, 2018AJ....156..178Z}.
However, the negative amplitude of Kepler 79 d is unlikely due to heterogeneity on the photosphere of its host star \citep{2020AJ....160..201C}. 

It is worth noting that the negative water amplitude can also originate from the atmospheric spectrum if the atmosphere is extremely hazy. 
For a very hazy exoplanet, a haze opacity dominates over gas opacity at both $1.25$ and $1.4 \mu$m.
Inserting Equation \eqref{eq:kappa_haze1} into Equation \eqref{eq:Ah2}, the water amplitude can be written by \footnote{Strictly speaking, Equation \eqref{eq:Ah2} is valid only for vertically constant opacity. We note that the vertical variation of haze opacity alters the gradient of the spectral slope but does not change its sign \citep{Ohno&Kawashima20}}
\begin{equation}\label{eq:Ah4}
    A_{\rm H}=\ln{\left( \frac{1.25}{1.4}\frac{f(1.4~{\rm {\mu}m})}{f(1.25~{\rm {\mu}m})} \right)}.
\end{equation}
Assuming that the optical constants (i.e., $f$ in Equation \ref{eq:kappa_haze2P}) are nearly invariant with wavelength from $\lambda=1.25$--$1.4~{\rm \mu m}$, which is true for the refractive indices of soot and Titan tholin \citep[see e.g., Figure 2 of][]{Lavvas&Aufaux21}, Equation \eqref{eq:Ah4} yields a negative amplitude of $A_{\rm H}\approx \ln{(1.25/1.4)}<0$.
In other words, the water amplitude becomes negative because the transit depth monotonically decreases with wavelength in the haze-induced spectral slope, i.e., $D(1.25~{\rm {\mu}m})>D(1.4~{\rm {\mu}m})$. The actual amplitude of the slope-induced $A_{\rm H}$ is heavily dependent on the optical properties of the haze particles (i.e., $f$ in Equation \eqref{eq:Ah4}), which are still largely unconstrained for the the exotic conditions in exoplanet atmospheres.

\section{Summary}\label{sec:summary}
In this study, we compiled and compared the haziness of 25 temperate to warm exoplanets using their transmission spectra data observed by \textit{HST}/WFC3. By examining the relationship between the water amplitude $A_{\rm H}$ and various planetary and stellar forcing parameters, we found some notable correlations among all the parameters we investigated. Our analysis shows that:

\begin{enumerate}
    \item Previously established linear trends between $A_{\rm H}$ vs. $T_{\rm eq}$ and $f_{\rm HHe}$ in \citetalias{2017AJ....154..261C} break down with the addition of new exoplanet data.
    \item Among all the stellar forcing parameters, orbital eccentricity and age of the star hold the best but tentative correlations with $A_{\rm H}$. Specifically, planets with more eccentric orbits with older parent stars tend to have clearer atmospheres.
    \item Among all the planetary parameters, planet gravity, atmospheric scale height, and planet density hold the best but tentative correlations with $A_{\rm H}$ (p $\leq$ 0.04). Specifically, planets with smaller scale heights (less puffy atmospheres), and larger bulk densities tend to be clearer.
    This potentially supports the idea that the water amplitudes are correlated with the planetary properties controlling the efficiency of haze removal, namely the scale height and gravity.
    \item Our simple analytical haze model showed that $A_{\rm H}$ may be dependent on a combination of planetary parameters (equilibrium temperature and planet gravity): $T^{1/2}/g$. This parameter holds a tentative trend with $A_{\rm H}$.
    However, its correlation with $A_{\rm H}$ is still not very good. This may indicate that our fixed parameters, such as haze mass flux and eddy diffusion coefficient, have implicit dependence on planetary parameters.
\end{enumerate}

In this study, we tentatively suggest that less hazy atmospheres exist for exoplanets with smaller atmospheric scale heights, higher surface gravity, bulk densities, orbital eccentricities, and with older stars. However, none of the parameters have statistically significant correlations with $A_{\rm H}$. This suggests that haziness in warm exoplanets may not be simply controlled by a single or a simple combination of planetary/stellar parameters. We still need more observations, laboratory experiments, and modeling work to fully understand the complex physical and chemical processes that lead to hazy exoplanet atmospheres. Note that as the field evolves with more upcoming transit observations, our interpretation might also change. Thus, we make a public-available website archiving all the data presented in this study: \url{https://exoplanethaziness.shinyapps.io/hazyweb/}. This would allow us to add any new observations and to keep track of the updated trends.

\section{Acknowledgements}
A. Dymont thanks the Other Worlds Laboratory at UC Santa Cruz for summer research support, Koret Foundation for the Koret Undergraduate Scholarship, and the UC Santa Cruz Undergraduate Research in Science and Technology award for additional funding. X. Yu is supported by the 51 Pegasi b Postdoctoral Fellowship from the Heising-Simons Foundation. K. Ohno is supported by the JSPS Overseas Research Fellowships. X. Zhang is supported by NASA Exoplanet Research Grant 80NSSC22K0236. X. Zhang and J. Fortney are supported by NASA Interdisciplinary Consortia for Astrobiology Research (ICAR) grant 80NSSC21K0597. We thank Dr. Kevin France for providing guidance on scaling stellar XUV fluxes. We thank Amaan Khwaja, Yash Rajpal, and Connor Dickinson for making the publicly-available website to archive the data presented in this paper.

\bibliography{references.bib}{}
\bibliographystyle{aasjournal}

\appendix 

\section{Extended Data Tables}\label{ap:A}

In this section, we categorize all studied parameters into three extended Data Tables. Table \ref{tab:s-param} contains all intrinsic star parameters. Table \ref{tab:sy-param} contains all other stellar forcing parameters. Table \ref{tab:p-param} contains all intrinsic and calculated planetary parameters.

\begin{longrotatetable}
\begin{deluxetable}{l|Cl|Cl|C@{$\pm$}Cl|C@{$\pm$}Cl|C@{$\pm$}Cl|C@{$\pm$}Cl|C@{$\pm$}Cl}
\tablecaption{\large Summary of Stellar Parameters\label{tab:s-param}}
\tablehead{
\colhead{Planet} & \multicolumn{2}{c}{Spectral} & \multicolumn{2}{c}{$t_{\rm age}$}  & \multicolumn{3}{c}{$M_*$ }  & \multicolumn{3}{c}{$R_*$ } & \multicolumn{3}{c}{$T_{\rm eff*}$} & \multicolumn{3}{c}{$Z_{[Fe/H]}$}&  \multicolumn{3}{c}{$P_{rot}$}  \\ \colhead{Name} & \multicolumn{2}{c}{Type} &\multicolumn{2}{c}{[Gyr]} & \multicolumn{3}{c}{[$M_{\odot}$]} & \multicolumn{3}{c}{[$R_{\odot}$]} & \multicolumn{3}{c}{[K]} & \multicolumn{3}{c}{[dex]} & \multicolumn{3}{c}{[day]}
}
\tabletypesize{\small}
\startdata
GJ 436 b   &M3    &[1]&  $6^{+4}_{-5}$    &[2] & 0.445     & 0.044     &[3]& 0.449  & 0.019  &[3]& 3479 & 60 &[2]&0.04 &0.17 &[4]& 44.6& 2.0&[5] \\[0.5ex]
GJ 1132 b  &M4.5  &[6]& 9$\pm$4    &[6] & 0.181     & 0.019     &[7]& 0.2105 & 0.0102 &[7]& 3270 & 140 &[7]&  -0.12 &    0.15 &[6]& 122.3  & 6.0 &[7] \\[0.5ex]
GJ 1214 b  &M4.5  &[1]& 6.5$\pm$3.5     &[9] & 0.176     & 0.0087    &[10]& 0.213  & 0.011  &[10]& 3252 & 20 &[10]&   0.05 &   0.09 &[10]& 125    & 5       &[11] \\[0.5ex]
GJ 3470 b  & M1.5 &[12]& 1.8$\pm$1.2   &[13] & 0.476     & 0.019     &[13]& 0.474  & 0.014  &[13]& 3725 & 54&[13]&   0.420 &   0.019 &[13]& 20.70   & 0.15    &[13] \\[0.5ex]
HAT-P-11 b &K4&[14]& $6.5^{+5.9}_{-4.1}$  &[14] & 0.81      & 0.027     &[14]& 0.683  & 0.009  &[15]& 4780 & 50 &[14]&   0.31 &   0.05 &[14]& 29.319 & 0.497   &[16]\\[0.5ex]
HAT-P-12 b &K4&[17]& 2.5$\pm$2      &[17] & 0.73      & 0.02      &[17]& 0.7    & 0.02   &[17]& 4650 & 60 &[17]&  -0.29 &   0.05 &[17]& 71     & 2       &[18]\tablenotemark{a}\\[0.5ex]
HAT-P-17 b &K&[19]& 7.8$\pm$3.3   &[20] & 0.92      & 0.15      &[21]& 0.84   & 0.04   &[21]& 5338 & 118 &[21]&   0.05 &   0.03 &[21]& 142    & 4       &[20]\tablenotemark{a}\\[0.5ex]
HAT-P-18 b &K2&[22]& $12.4^{+4.4}_{-6.4}$  &[22] & 0.77      & 0.03      &[22]& 0.75   & 0.04   &[22]& 4803 & 80 &[22]&   0.10 &   0.08 &[22]& 76     & 4       &[20]\tablenotemark{a}\\[0.5ex]
HAT-P-26 b &K1&[23]& $9^{+3}_{-4.9}$   &[23] & 0.816     & 0.033     &[24]& 0.788  & 0.098  &[24]& 5079 & 88 &[24]&  -0.04 &   0.08 &[24]& 48     & 4       &[23]  \\[0.5ex]
HAT-P-38 b  &G5&[25]& $10.1^{+3.9}_{-4.8}$   &[20] & 0.89      & 0.04      &[26]& 0.92   & 0.1    &[26]& 5330 & 100&[26]&   0.06 &    0.1 &[20]& 116    & 10.1    &[20]\tablenotemark{a}\\[0.5ex]
HD 3167 c  &K0V&[27]& 7.8$\pm$4.3   &[28] & 0.866     & 0.0333    &[27]& 0.872  & 0.057  &[27]& 5261 & 60 &[27]&   0.04 &   0.05 &[27]& 23.52  & 2.87    &[27] \\[0.5ex]
HD 97658 b &K1V&[29]& 6.1$\pm$0.7  &[30] & 0.77      & 0.05      &[31]& 0.746  & 0.034  &[32]& 5192 & 122 &[32]&  -0.23 &   0.03 &[31]& 34     & 2       &[33] \\[0.5ex]
HD 106315 c&F5&[34]& 4.48$\pm$0.96 &[35] & 1.091     & 0.036     &[35]& 1.296  & 0.058  &[35]& 6327 & 48 &[35]&  -0.31 &   0.08 &[35]& 5.15   & 0.28    &[35] \\[0.5ex]
HIP 41378 f& F6&[36]& 3.1^{+0.4}_{-0.6}   &[37] & 1.16      & 0.04      &[37]& 1.27   & 0.01   &[37]& 6321 & 48.00 &[37]&  -0.09 &   0.07 &[37]& 6.4    & 0.8     &[37] \\[0.5ex]
Kepler-51 b&G&[38]& 0.5 $\pm$ 0.25  &[38] & 0.985     & 0.012     &[38]& 0.881  & 0.011  &[38]& 5662 & 65&[39]&   0.04 &   0.04 &[39]& 8.222  & 0.007   &[40] \\[0.5ex]
Kepler-51 d&G&[38]& 0.5 $\pm$ 0.25  &[38] & 0.985     & 0.012     &[38]& 0.881  & 0.011  &[38]& 5662 & 65&[39]&   0.04 &   0.04 &[39]& 8.222  & 0.007   &[40] \\[0.5ex]
Kepler-79 d &F&[41]& $1.3^{+1}_{-0.4}$&[41] & 1.244     & 0.042     &[42]& 1.316  & 0.038  &[42]& 6389 & 60 &[42]&   0.06 &   0.04 &[42]& 2.4    & 2.4     &[41]\\[0.5ex]
K2-18 b  &M3&[43] &2.4  $\pm$  0.6   &[44] & 0.4951    & 0.0043    &[45]& 0.4445 & 0.0148 &[43]& 3457 & 39 &[46]&   0.12 &    0.2 &[46]& 39.63  & 0.5     &[47]  \\[0.5ex]
LHS 1140 b &M4.5&[48]&9    $\pm$ 4     &[48] & 0.179     & 0.014     &[49]& 0.2139 & 0.0041 &[49]& 3216 & 39 &[49]&  -0.24 &    0.1 &[48]& 131    & 5       &[48]  \\[0.5ex]
TOI-674 b &M2 &[50]& $5.5^{+2.9}_{1.9}$   &[50] & 0.42      & 0.01      &[50]& 0.42   & 0.013  &[50]& 3514 & 57 &[50]&   0.17 &    0.1 &[50]& 52     & 5       &[50]\\[0.5ex]
WASP-29 b  &K4&[51] &10.5 $\pm$ 3.5   &[20] & 0.77      & 0.25      &[24]& 0.79   & 0.07   &[24]& 4800 & 150 &[24]&   0.11 &    0.14 &[52]& 27     & 5.9     &[20]\tablenotemark{a}\\[0.5ex]
WASP-67 b  &K0&[24]&$12.6^{+1}_{-4.2}$&[20] & 0.91      & 0.28      &[24]& 0.88   & 0.08   &[24]& 5200 & 100 &[24]&  -0.07 &   0.09 &[24]& 21     & 10.1    &[20]\tablenotemark{a}\\[0.5ex]
WASP-69 b &K5&[53] &1    $\pm$ 1     &[53] & 0.98      & 0.14      &[20]& 0.86   & 0.03   &[20]& 4700 & 50 &[20]&   0.15 &   0.08 &[53]& 23.07  & 0.16    &[20]\\[0.5ex]
WASP-80 b  &K7&[54] &7  $\pm$ 7     &[20] & 0.58      & 0.05      &[55]& 0.59   & 0.02   &[55]& 4143 & 92 &[55]&  -0.13 &    0.1 &[55]& 8.5    & 0.8     &[56]\\[0.5ex]
WASP-107 b &K6&[57]&8.3 $\pm$ 4.3   &[58] & 0.683     & 0.017     &[59]& 0.67   & 0.02   &[59]& 4425 & 70&[59]&   0.02 &   0.09 &[59]& 17.1   & 1.0      &[57]\\[0.5ex]
\enddata
\tablenotetext{a}{ We calculate the lower bound stellar rotation period from $v\sin(i)$ measurement}
\tablerefs{
[1] \citet{2013ApJ...763..149F}; [2] \citet{2007ApJ...671L..65T}; [3] \citet{2015ApJ...804...64M}; [4] \citet{2012ApJ...748...93R}; [5] \citet{2019AA...621A.126D};
[6] \citet{2015Natur.527..204B}; [7] \citet{2018AA...618A.142B};
[9] \citet{2009Natur.462..891C}; [10] \citet{2013AA...551A..48A}; [11] \cite{2018AA...614A..35M};
[12] \citet{2019AJ....157...97K}; [13] \citet{2020AA...638A..61P};
[14] \citet{2010ApJ...710.1724B}; [15] \citet{2011ApJ...740...33D}; [16] \citet{2015ApJ...801....3M};
[17] \citet{2009ApJ...706..785H}; [18] \citet{2018AA...613A..41M};
[19] \citet{2012ApJ...749..134H}; [20] \citet{2017AA...602A.107B}; [21] \citet{2019AJ....158..138S}
[22] \citet{2011ApJ...726...52H};
[23] \citet{2011ApJ...728..138H}; [24] \citet{2017AJ....153..136S};
[25] \citet{2018yCat.5153....0L}; [26] \citet{2012PASJ...64...97S};
[27] \citet{2021AJ....161...18M}; [28] \citet{2017AJ....154..122C};
[29] \citet{2003AJ....126.2048G}; [30] \citet{2011arXiv1109.2549H}; [31] \citet{2014ApJ...786....2V}; [32] \citet{2018yCat.1345....0G}; [33]\citet{2020AJ....159..239G};
[34] \citet{2020arXiv200607444K}; [35] \citet{2017AA...608A..25B}
[36] \citet{2018yCat.5153....0L}; [37] \citet{2019arXiv191107355S}
[38] \citet{2020AJ....159...57L}; [39] \citet{2017AJ....154..108J}; [40] \citet{2013ApJ...775L..11M}; 
[41] \citet{2020AJ....160..201C}; [42] \citet{2018AJ....156..264F};
[43] \citet{2019ApJ...887L..14B}; [44] \citet{2019RNAAS...3..189G}; [45] \citet{2019AA...621A..49C}; [46] \citet{2017ApJ...834..187B}; [47] \citet{2018AJ....155..257S};
[48] \citet{2017Natur.544..333D}; [49] \citet{2019AJ....157...32M};
[50] \citet[][]{2021AA...653A..60M}
[51] \citet{2011AA...529A.136E}; [52] \citet{2010ApJ...723L..60H}
[53] \citet{2014MNRAS.445.1114A};
[54] \citet{2015AA...576A..42S}; [55] \citet{2015MNRAS.450.2279T} [56] \citet{2013AA...551A..80T};
[57] \citet{2017AA...604A.110A}; [58] \citet{2017MNRAS.469.1622M}; [59] \citet{2021AJ....161...70P}
}
\end{deluxetable}
\end{longrotatetable}

\begin{longrotatetable}
\begin{deluxetable}{l|C@{$\pm$}Cl|C@{$\pm$}Cl|C@{$\pm$}C|C@{$\pm$}C|R@{$\pm$}R|R|R|R}
\tablecaption{\large Summary of Stellar Forcing Parameters\label{tab:sy-param}}
\tablehead{
\colhead{Planet}& \multicolumn{3}{c}{$a$} &  \multicolumn{3}{c}{$e$}  & \multicolumn{2}{c}{$\rho_*$} & \multicolumn{2}{c}{$g_*$} & \multicolumn{2}{c}{$\log(L_{\rm bol})$\tablenotemark{a}} & \multicolumn{1}{c}{$F_{\rm NUV}$} & \multicolumn{1}{c}{$F_{\rm FUV}$} & \multicolumn{1}{c}{$F_{\rm XUV}$}  \\ \colhead{Name} & \multicolumn{3}{c}{[AU]} & \multicolumn{3}{c}{} & \multicolumn{2}{c}{[$\rho_{\odot}$]} & \multicolumn{2}{c}{[$m^-2$]} & \multicolumn{2}{c}{[dex(erg s$^{-1}$)]} & \multicolumn{1}{c}{\tiny[erg cm$^{-2}$ s$^{-1}$]} & \multicolumn{1}{c}{\tiny[erg cm$^{-2}$ s$^{-1}$]} & \multicolumn{1}{c}{\tiny[erg cm$^{-2}$ s$^{-1}$]}}
\decimals
\startdata
GJ 436 b   & 0.0308 & 0.0013 &[1]& 0.162 & 0.0041&[1]& 4.92 & 0.79& 605 & 79 & 32.01 & 0.05 & 3650 & 1150 & 1490.00  \\[0.5ex]
GJ 1132 b  & 0.0153 & 0.0005 &[2]& 0.00 & 0.00   &[3]& 19.41 & 3.48& 1119 & 160 & 31.24 & 0.09 & 2590 & 983 & 4090  \\[0.5ex]
GJ 1214 b  & 0.0148 & 0.0008 &[4]& 0.000 & 0.270 &[5]& 18.21 & 2.96& 1063 & 122 & 31.24 & 0.05 & 1800 & 1010 & 900  \\[0.5ex]
GJ 3470 b  & 0.0285 & 0.0018 &[6]& 0.114 & 0.052 &[7]& 4.47 & 0.43& 581 & 41 & 32.17 & 0.04 & 64700 & 13300 & 12200  \\[0.5ex]
HAT-P-11 b & 0.055  & 0.001  &[8]& 0.265 & 0.0007&[9]& 2.54 & 0.13& 476 & 20 & 32.92 & 0.02 & 634000 & 23800 & 9290 \\[0.5ex]
HAT-P-12 b & 0.038  & 0.001 &[10]& 0.00 & 0.00  &[10]& 2.13 & 0.19& 408 & 26 & 32.90 & 0.03 & 398000 & 6600 & 2180  \\[0.5ex]
HAT-P-17 b & 0.089  & 0.005 &[11]& 0.35 & 0.01  &[11]& 1.55 & 0.34& 357 & 68 & 33.30 & 0.06 & 850000 & 2970 & 568  \\[0.5ex]
HAT-P-18 b & 0.056  & 0.004 &[12]& 0.08 & 0.05  &[12]& 1.82 & 0.30& 375 & 43 & 33.01 & 0.05 & 342000 & 4010 & 667  \\[0.5ex]
HAT-P-26 b & 0.0442 & 0.0055&[13]& 0.12 & 0.06  &[13]& 1.67 & 0.63& 360 & 91 & 33.15 & 0.11 & 1400000 & 8770 & 1600  \\[0.5ex]
HAT-P-38 b & 0.052  & 0.007 &[14]& 0.067& 0.047 &[14]& 1.14 & 0.38& 288 & 64 & 33.37 & 0.10 & 2900000 & 10300 & 574  \\[0.5ex]
HD 3167 c  & 0.1795	&0.0023 &[15]& 0.267 & 0.00 &[16]& 1.31 & 0.26& 312 & 43 & 33.30 & 0.06 & 179000 & 742 & 484  \\[0.5ex]
HD 97658 b & 0.093  & 0.004 &[17]& 0.030 & 0.034&[17]& 1.85 & 0.28& 379 & 42 & 33.14 & 0.06 & 515000 & 2520 & 1420  \\[0.5ex]
HD 106315 c& 0.1538 & 0.0100&[18]& 0.22 & 0.15  &[18]& 0.50 & 0.07& 178 & 17 & 33.97 & 0.04 & 12000000 & 4430 & 2000  \\[0.5ex]
HIP 41378 f& 1.36   & 0.01  &[19]& 0.01 & 0.01  &[19]& 0.57 & 0.02& 197.09 & 7.47 & 33.95 & 0.01 & 143000 & 53.70 & 10.40  \\[0.5ex]
Kepler-51 b& 0.24   & 0.01  &[20]& 0.04 & 0.01  &[20]& 1.44 & 0.06& 348 & 10 & 33.44 & 0.02 & 330000 & 552 & 1030  \\[0.5ex]
Kepler-51 d& 0.51   & 0.02  &[20]& 0.01 & 0.01  &[20]& 1.44 & 0.06& 348 & 10 & 33.44 & 0.02 & 73800 & 123 & 230  \\[0.5ex]
Kepler-79 d& 0.2937 & 0.0027&[21]& 0.0250 & 0.0590&[22]& 0.55 & 0.05& 197 & 13 & 34.00 & 0.03 & 4040000 & 1300 & 616  \\[0.5ex]
K2-18 b    & 0.15910& 0.00047&[23]&0.20 & 0.08  &[24]& 5.64 & 0.56& 687 & 46 & 31.99 & 0.03 & 202 & 50.0 & 77.5  \\[0.5ex]
LHS 1140 b & 0.0936 & 0.0024&[25]& 0.00 & 0.06  &[25]& 18.29 & 1.78& 1072 & 93 & 31.23 & 0.03 & 59.3 & 25.4 & 104  \\[0.5ex]
TOI-674 b  & 0.025 & 0.0008 &[26]& 0.00 & 0.00  &[26]& 5.67 & 0.54& 652.48 & 43.28 & 31.97 & 0.04 & 8850.00 & 1930.00 & 2180.00  \\[0.5ex]
WASP-29 b  & 0.045  & 0.005 &[27]& 0.03 & 0.04  &[27]& 1.56 & 0.66& 338 & 125 & 33.06 & 0.09 & 588000 & 6950 & 1060  \\[0.5ex]
WASP-67 b  & 0.053  & 0.005 &[27]& 0.00 & 0.00  &[27]& 1.33 & 0.55& 322 & 115 & 33.29 & 0.09 & 1770000 & 8450 & 5420  \\[0.5ex]
WASP-69 b  & 0.048  & 0.002 &[27]& 0.00 & 0.00  &[27]& 1.54 & 0.27& 363 & 58 & 33.10 & 0.04 & 1320000 & 49800 & 19400  \\[0.5ex]
WASP-80 b  & 0.035  & 0.001 &[28]& 0.002 & 0.010 &[28]& 2.82 & 0.38& 457 & 50 & 32.55 & 0.05 & 70800 & 3710 & 9230  \\[0.5ex]
WASP-107 b & 0.0558 & 0.0018&[29]& 0.06 & 0.04  &[29]& 2.27 & 0.21& 417 & 26 & 32.77 & 0.04 & 85700 & 2370 & 3260  \\[0.5ex]
\enddata
\tablenotetext{a}{$\log_{10}(L_{\rm bol}) = \log_{10}(\sigma_{SB}T_{\rm eff*}^44{\pi}R_*^2) $, where $\sigma_{SB}$ is the Stephan-Boltzmann constant.}
\tablerefs{
[1] \citet{2014AA...572A..73L};
[2] \citet{2018AA...618A.142B}; [3] \citet{2017AJ....153....9C};
[4] \citet{2013AA...551A..48A}; [5] \citet{2009Natur.462..891C};
[6] \citet{2015ApJ...814..102D}; [7] \citet{2019AJ....157...97K};
[8] \citet{2019AJ....158..244C}; [9] \citet{2017AA...597A.113H};
[10] \citet{2009ApJ...706..785H};
[11] \citet{2017AJ....153..136S};
[12] \citet{2011ApJ...726...52H};
[13] \citet{2011ApJ...728..138H};
[14] \citet{2012PASJ...64...97S};
[15] \citet{2021AJ....161...18M}; [16] \citet{2017AJ....154..122C};
[17] \citet{2020AJ....159..239G};
[18] \citet{2017AA...608A..25B};
[19] \citet{2019arXiv191107355S}
[20] \citet{2020AJ....159...57L};
[21] \citet{2018AJ....156..264F}; [22] \citet{2014ApJ...785...15J};
[23] \citet{2019ApJ...887L..14B}; [24] \citet{2018AJ....155..257S};
[25] \citet{2019AJ....157...32M};
[26] \citet{2021AA...653A..60M}
[27] \citet{2017AJ....153..136S};
[28] \citet{2015MNRAS.450.2279T};
[29] \citet{2017AJ....153..205D};
}
\end{deluxetable}
\end{longrotatetable}

\begin{longrotatetable}
\begin{deluxetable}{l|C@{$\pm$}Cl|C@{$\pm$}Cl|C@{$\pm$}C|C@{$\pm$}C|C@{$\pm$}C|C@{$\pm$}C|C@{$\pm$}C}
\tablecaption{\large{Summary of Planetary Parameters}\label{tab:p-param}}
\tablehead{
\colhead{Planet} & \multicolumn{3}{c}{$M_{\rm p}$}  & \multicolumn{3}{c}{$R_{p}$} & \multicolumn{2}{c}{$f_{\rm HHe}$} & \multicolumn{2}{c}{$T_{\rm eq}$} & \multicolumn{2}{c}{$\rho_{\rm p}$} & \multicolumn{2}{c}{$g_{\rm p}$} & \multicolumn{2}{c}{$H$}  \\ \colhead{Name} & \multicolumn{3}{c}{[$M_{\oplus}$]} & \multicolumn{3}{c}{[$R_{\oplus}$]} & \multicolumn{2}{c}{[\%]} & \multicolumn{2}{c}{[K]} & \multicolumn{2}{c}{$\rho_{\oplus}$} & \multicolumn{2}{c}{[m$^2$/s]} & \multicolumn{2}{c}{[km]}
}
\decimals
\startdata 
GJ 436 b    & 25.4  & 2.1  &[1]  & 4.10   & 0.16 &[1]  & 12.87 & 1.70 & 586       & 10        & 0.369  & 0.0528 & 14.81 & 1.68 & 143.00 &   16.4  \\[0.5ex]
GJ 1132 b   & 1.66  & 0.23 &[2]  & 1.13   & 0.056&[3]  & 0.01 & 0.00 & 535       & 23        & 1.15   & 0.234  & 12.74 & 2.17 & 152.00 &   26.7  \\[0.5ex]
GJ 1214 b   & 6.26  & 0.91 &[4]  & 2.80   & 0.24 &[4]  & 3.83 & 7.13 & 544       & 3         & 0.285  & 0.0842 & 7.82 & 1.76 & 252.00 &   56.6  \\[0.5ex]
GJ 3470 b   & 12.58 & 1.31 &[5]  & 3.88   & 0.32 &[6]  & 12.80 & 5.15 & 67        & 10        & 0.215  & 0.0578 & 8.19 & 1.60 & 296.00 &   57.9  \\[0.5ex]
HAT-P-11 b  & 23.4  & 1.5  &[7]  & 4.36   & 0.06 &[7]  & 15.10 & 2.57 & 74        & 8         & 0.282  & 0.0215 & 12.06 & 0.84 & 222.00 &   15.6  \\[0.5ex]
HAT-P-12 b  & 67.1  & 3.8  &[8]  & 10.7   & 0.3  &[8]  & 80.30 & 4.01 & 877       & 11        & 0.054  & 0.00578 & 5.69 & 0.47 & 557.00 &   46.8  \\[0.5ex]
HAT-P-17 b  & 184   & 19.0 &[9]  & 11.8   & 0.4  &[9]  & 90.00 & 4.00 & 724       & 16        & 0.113  & 0.0174 & 13.04 & 1.67 & 201.00 &   26.2  \\[0.5ex]
HAT-P-18 b  & 62.6  & 4.1  &[10] & 11.2   & 0.6  &[10] & 87.10 & 13.30 & 776       & 13        & 0.045  & 0.00768 & 4.93 & 0.61 & 569.00 &   70.9  \\[0.5ex]
HAT-P-26 b  & 18.75 & 2.23 &[11] & 6.03   & 0.75 &[12] & 31.70 & 6.20 & 946       & 16        & 0.086  & 0.0335 & 5.06 & 1.39 & 677.00 &  186.8  \\[0.5ex]
HAT-P-38 b  & 84.9  & 6.4  &[13] & 9.247  & 1.031&[13] & 62.80 & 11.90 & 988       & 19        & 0.107  & 0.0368 & 9.72 & 2.29 & 367.00 &   86.7  \\[0.5ex]
HD 3167 c   & 9.8   & 1.3  &[14] & 3.01   & 0.42 &[14] & 5.20 & 3.36 & 512       & 6         & 0.359  & 0.158  & 10.6 & 3.27 & 174.00 &   54.0  \\[0.5ex]
HD 97658 b  & 7.81  & 0.55 &[15] & 2.303  & 0.110&[15] & 0.99 & 1.80 & 65        & 15        & 0.639  & 0.102  & 14.43 & 1.71 & 163.00 &   19.7  \\[0.5ex]
HD 106315 c & 15.2  & 3.7  &[16] & 4.35   & 0.23 &[16] & 13.96 & 2.03 & 81        & 6         & 0.185  & 0.0537 & 7.87 & 2.09 & 372.00 &   98.8  \\[0.5ex]
HIP 41378 f & 12.3  & 3.1  &[17] & 9.2    & 0.1  & [17] & 71.00 & 4.00 & 269       & 2         & 0.016  & 0.00401 & 1.42 & 0.36 & 683.00 &  172.8  \\[0.5ex]
Kepler-51 b & 3.69  & 1.86 &[18] & 6.89   & 0.14 &[18] & 17.50 & 5.20 & 477       & 5         & 0.011  & 0.00573 & 0.76 & 0.39 & 2270.00 & 1146.0  \\[0.5ex]
Kepler-51 d & 5.70  & 1.12 &[18] & 9.46   & 0.16 &[18] & 35.20 & 7.60 & 328       & 4         & 0.007  & 0.00137 & 0.62 & 0.12 & 1900.00 &  379.7  \\[0.5ex]
Kepler-79 d & 5.3   & 0.9  &[19] & 7.2    & 0.2  &[19] & 36.70 & 3.56 & 597       & 6         & 0.014  & 0.00275 & 1.02 & 0.18 & 2120.00 &  380.1  \\[0.5ex]
K2-18 b     & 8.63  & 1.35 &[20] & 2.610  & 0.087&[21] & 2.82 & 0.52 & 255       & 3         & 0.485  & 0.0901 & 12.41 & 2.11 &  74.20 &   12.7  \\[0.5ex]
LHS 1140 b  & 6.98  & 0.89 &[22] & 1.727  & 0.032&[22] & 0.06 & 0.05 & 214       & 3         & 1.355  & 0.188  & 22.93 & 3.04 &  33.80 &    4.5  \\[0.5ex]
TOI-674 b   & 23.6  & 3.3  &[23] & 5.25   & 0.17 &[23] & 32.00 & 2.00 & 635       & 10        & 0.163  & 0.0278 & 8.39 & 1.29 & 274.00 &   42.4  \\[0.5ex]
WASP-29 b   & 73    & 16   &[9]  & 8.6    & 0.8  &[9]  & 60.00 & 7.57 & 89        & 28        & 0.114  & 0.0397 & 9.61 & 2.73 & 334.00 &   95.4  \\[0.5ex]
WASP-67 b   & 137   & 29   &[9]  & 12.9   & 1.2  &[9]  & 85.00 & 7.00 & 939       & 18        & 0.064  & 0.0227 & 8.06 & 2.29 & 421.00 &  119.7  \\[0.5ex]
WASP-69 b   & 92    & 10   &[9]  & 12.4   & 0.4  &[9]  & 88.00 & 5.00 & 878       & 9         & 0.048  & 0.00716 & 5.83 & 0.74 & 544.00 &   68.8  \\[0.5ex]
WASP-80 b   & 171   & 11   &[9]  & 11.2   & 0.3  &[9]  & 88.00 & 4.00 & 754       & 17        & 0.122  & 0.0135 & 13.36 & 1.18 & 204.00 &   18.6  \\[0.5ex]
WASP-107 b  & 30.5  & 1.7  &[24] & 10.39  & 0.33 &[25] & 84.00 & 2.00 & 676       & 11        & 0.027  & 0.00299 & 2.77 & 0.23 & 884.00 &   75.8  \\[0.5ex]
\enddata
\tablerefs{
[1] \citet{2014AA...572A..73L}
[2] \citet{2018AA...618A.142B}; [3] \citet{2017AJ....154..142D}; 
[4] \citet{2013AA...551A..48A};
[5] \citet{2019AJ....157...97K}; [6] \citet{2014MNRAS.443.1810B};
[7] \citet{2018AJ....155..255Y};
[8] \citet{2009ApJ...706..785H}; 
[9] \citet{2017AJ....153..136S};
[10] \citet{2011ApJ...726...52H};
[11] \citet{2011ApJ...728..138H}; [12] \citet{2019AA...628A.116V};
[13] \citet{2012PASJ...64...97S}; 
[14] \citet{2017AJ....154..122C};
[15] \citet{2020AJ....159..239G}
[16] \citet{2017AA...608A..25B};
[17] \citet{2019arXiv191107355S}
[18] \citet{2020AJ....159...57L};
[19] \citet{2020AJ....160..201C}; 
[20] \citet{2019AA...621A..49C}; [21] \citet{2019ApJ...887L..14B};
[22] \citet{2019AJ....157...32M};
[23] \citet[][]{2021AA...653A..60M}
[24] \citet{2021AJ....161...70P}; [25] \citet{2017AJ....153..205D}
}
\end{deluxetable}
\end{longrotatetable}

\section{Extended Figures\label{ap:B}}

\begin{figure}[h!]
    \centering
    \includegraphics[scale=0.29]{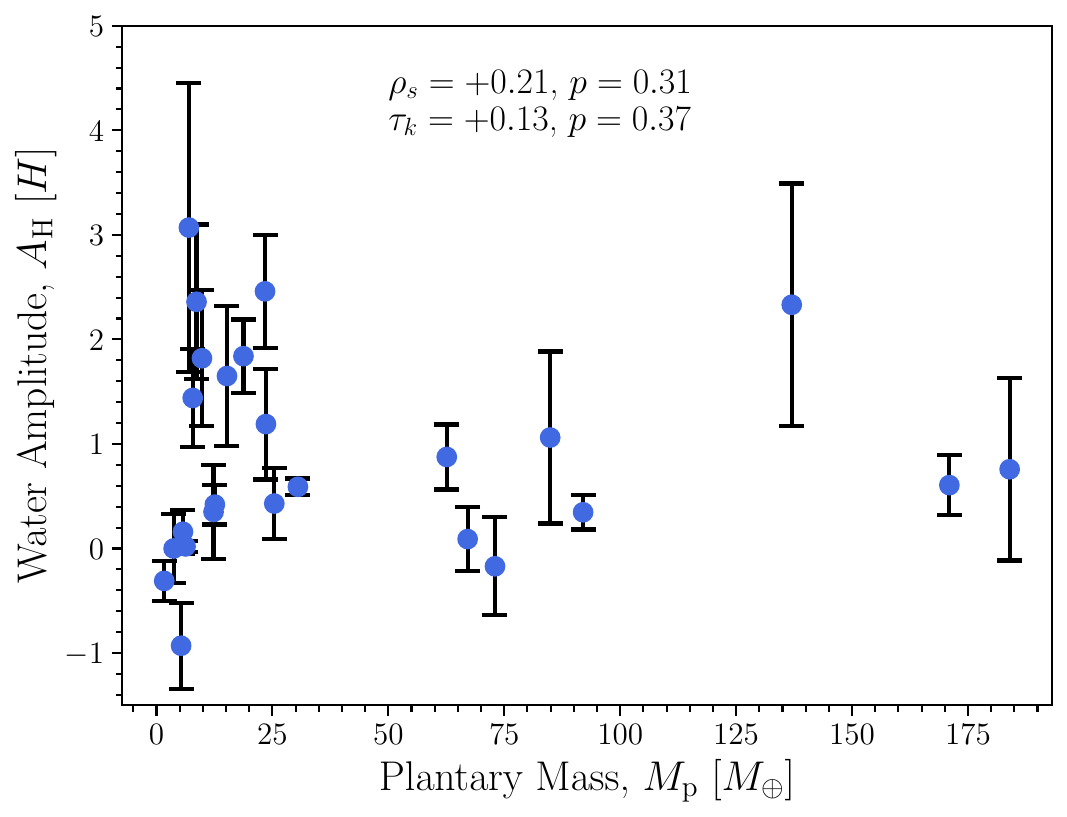}
    \includegraphics[scale=0.29]{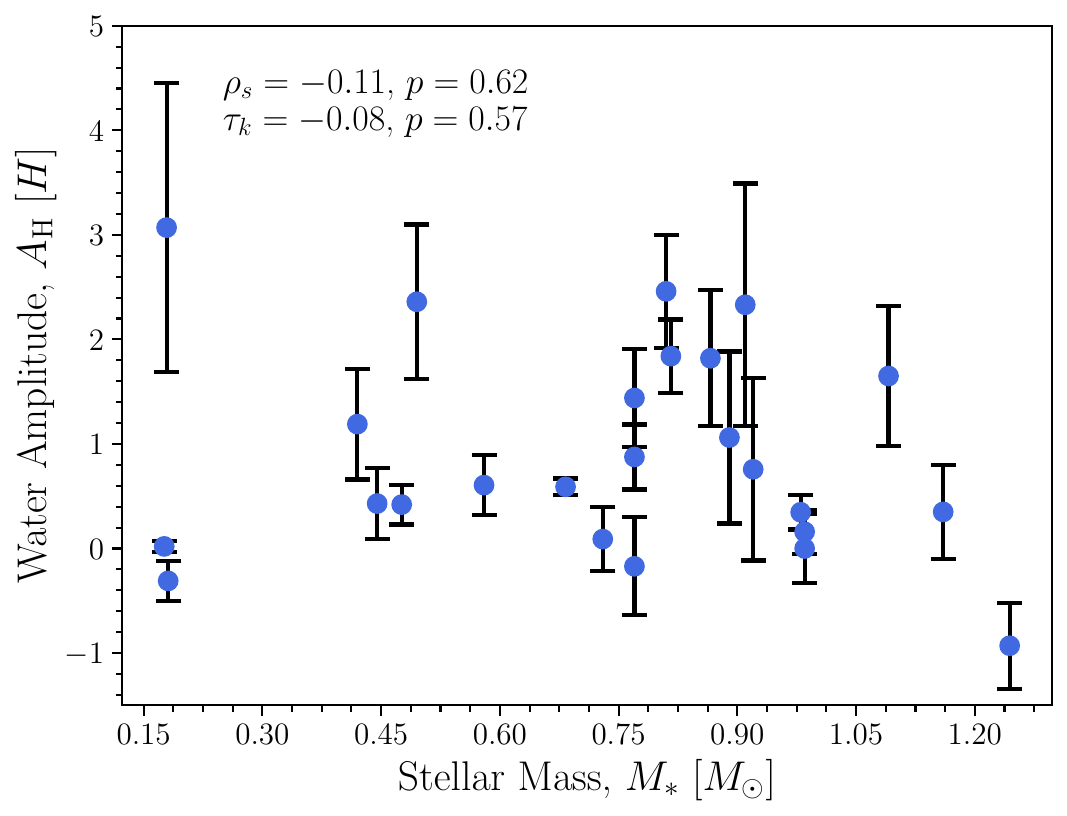}
    \includegraphics[scale=0.29]{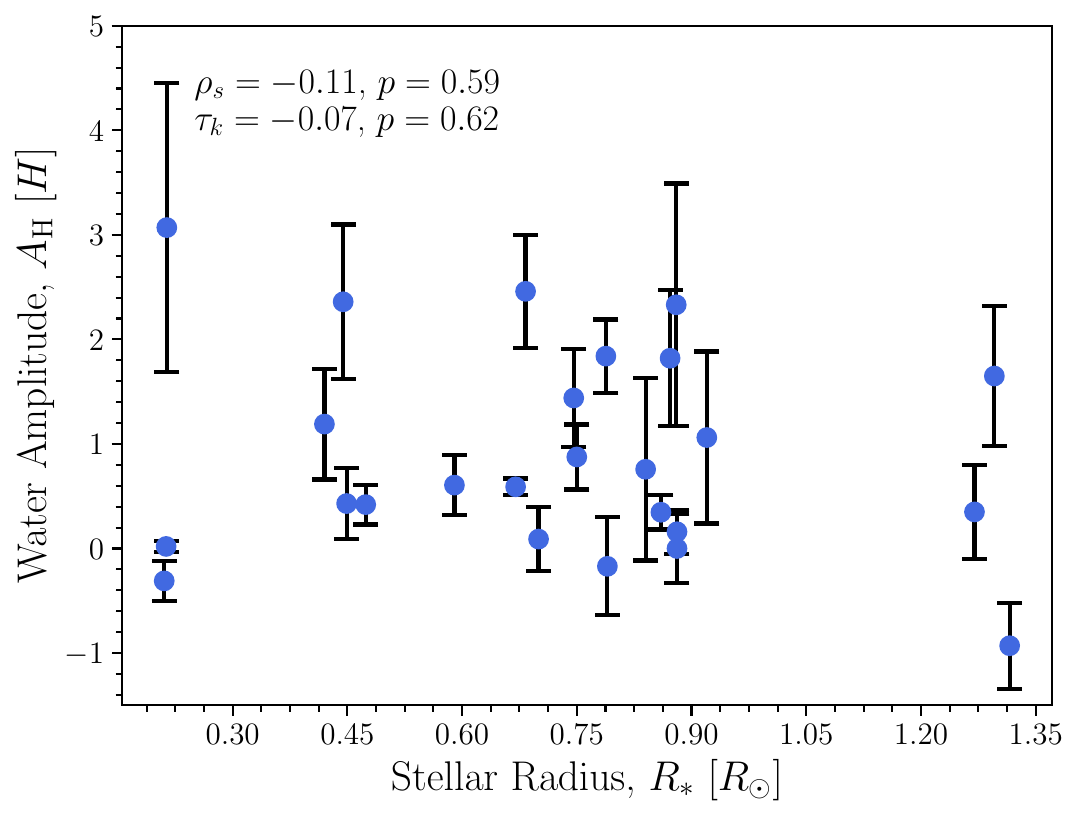}
    \includegraphics[scale=0.29]{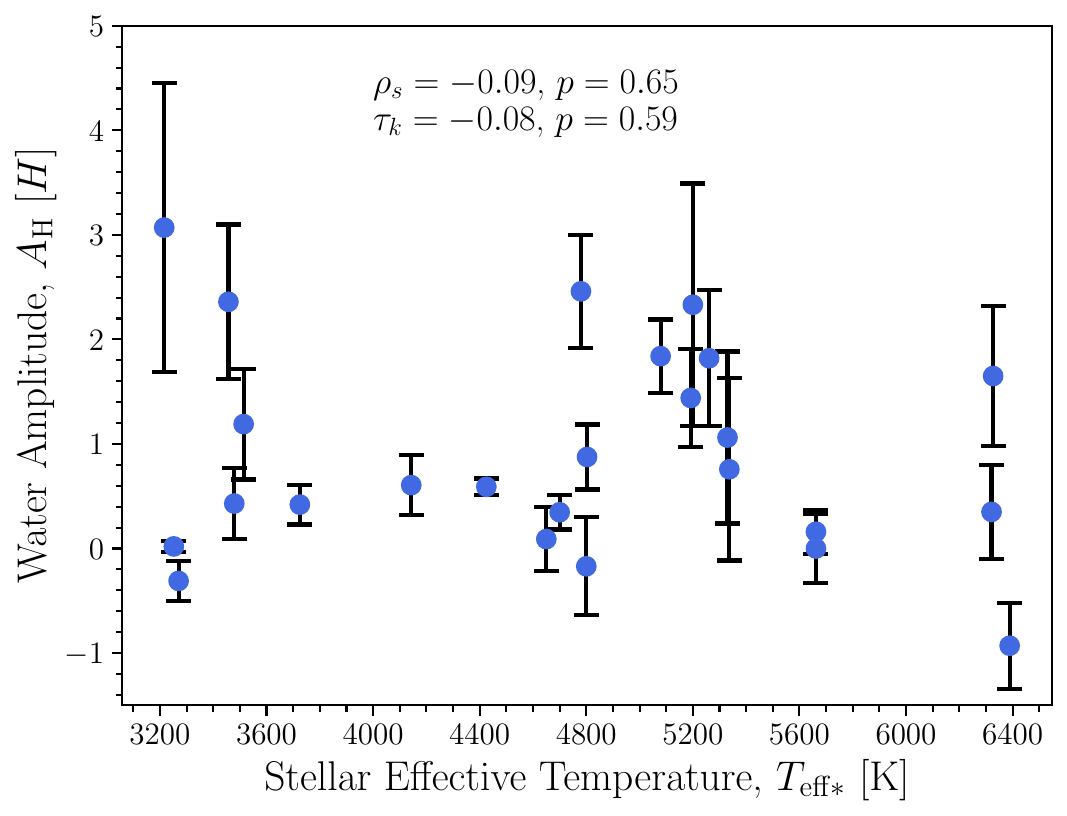}
    \includegraphics[scale=0.29]{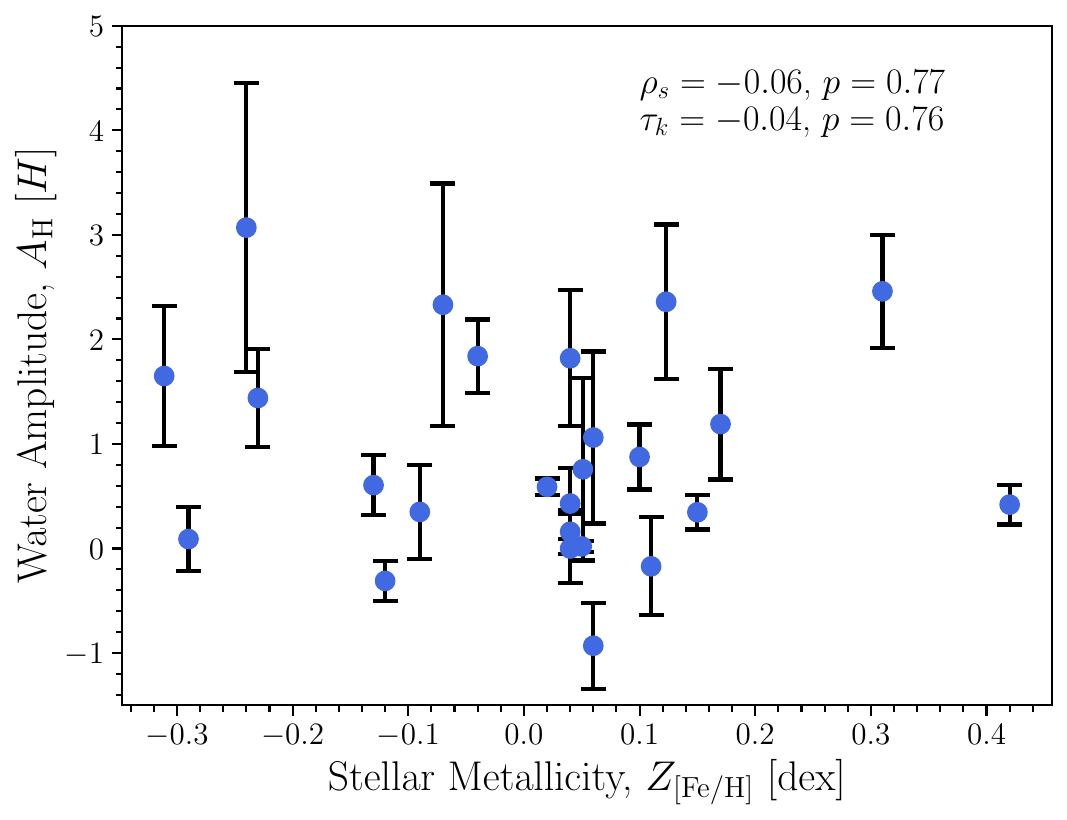}
    \includegraphics[scale=0.29]{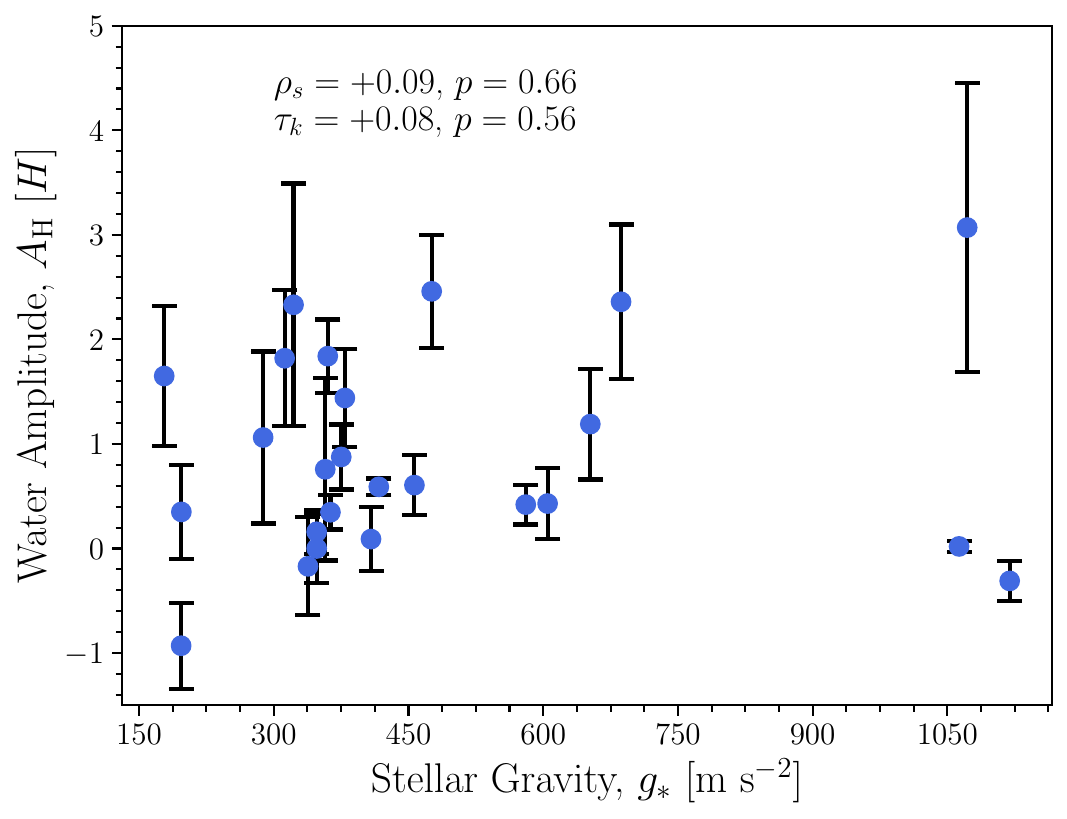}
    \includegraphics[scale=0.29]{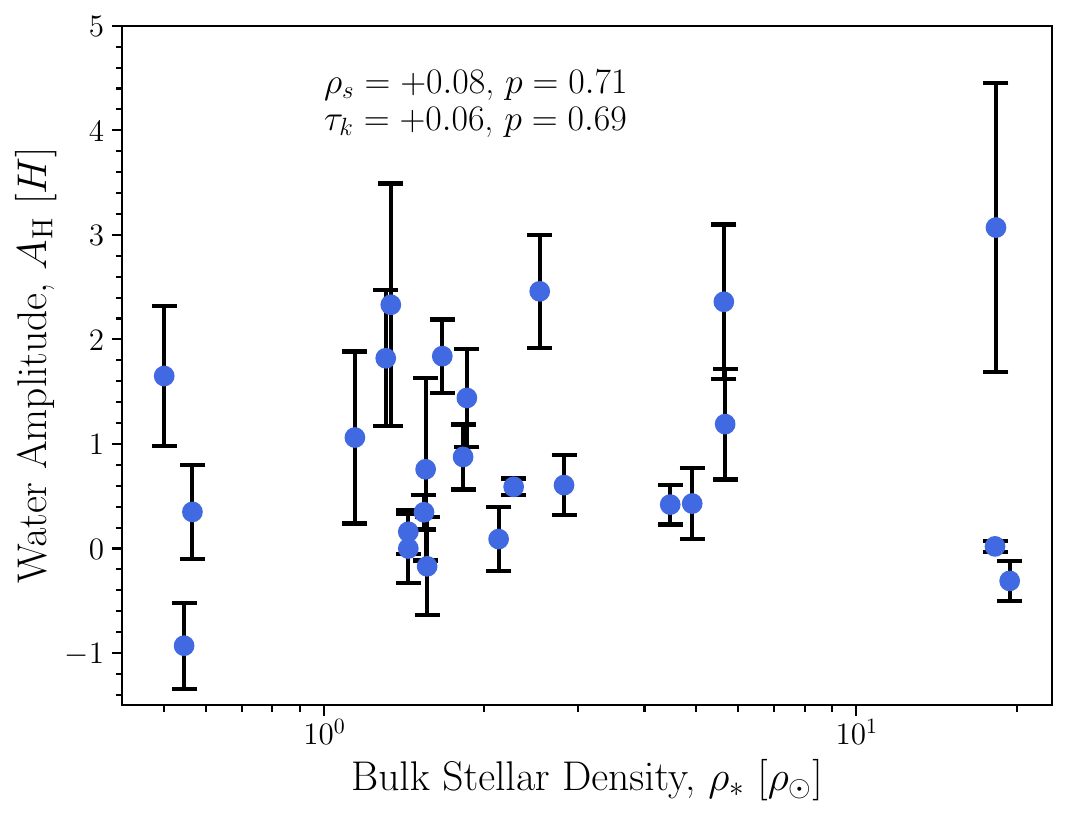}
    \includegraphics[scale=0.29]{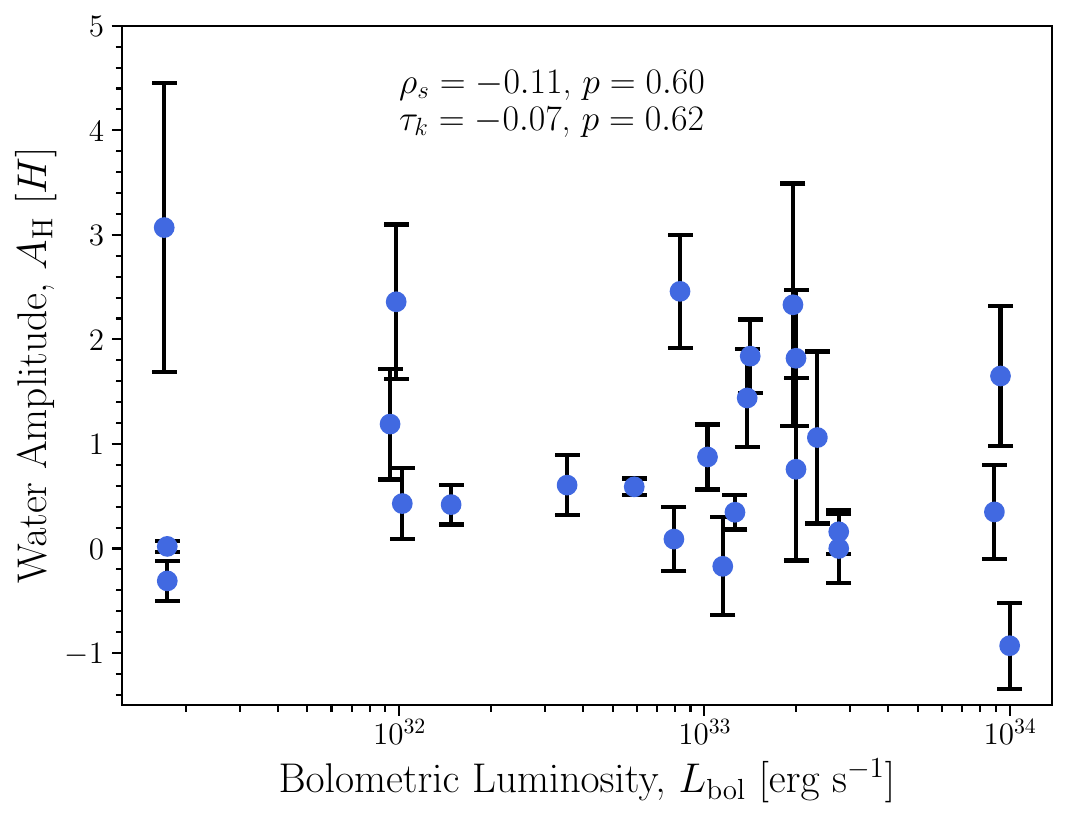}
    \includegraphics[scale=0.29]{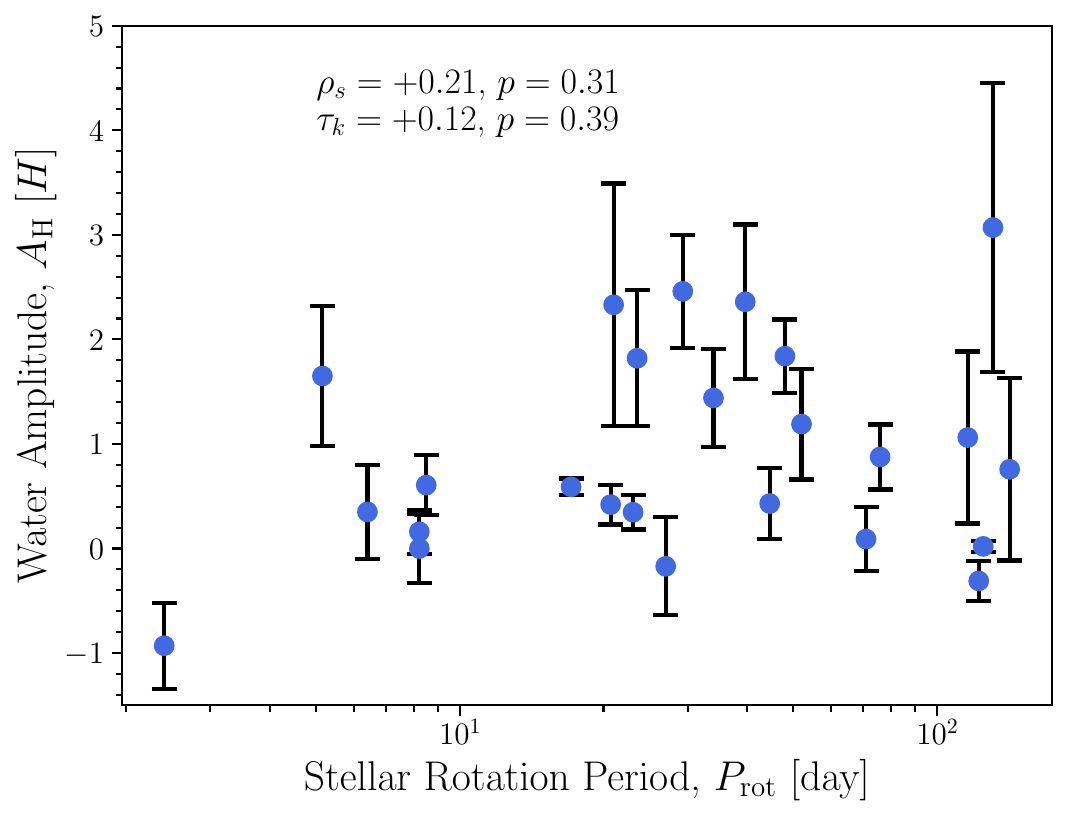}
    \caption{Contains all trends we compared of $A_{\rm H}$ vs parameters that were not discussed individually in the final analysis or discussion due to poor statistic (all). The vertical error bars (black) represent the estimated uncertainty of $A_{\rm H}$.}
    \label{fig:extra}
\end{figure}

\end{document}